\documentclass[preprint,5p,times,twocolumn,authoryear,sort]{elsarticle}
\pdfoutput=1
\usepackage{helvet}
\usepackage[T1]{fontenc}
\usepackage{amsmath,amsfonts,amssymb,mathrsfs,gensymb}
\usepackage{latexsym}
\usepackage{graphics}
\usepackage{graphicx}
\usepackage{hyperref}
\usepackage{epigraph}
\usepackage{epsf}
\usepackage{aniszewski}
\usepackage{color}
\usepackage[draft,textsize=small,color=green]{todonotes} 
\usepackage[yyyymmdd,hhmmss]{datetime} 
\usepackage{siunitx}

\title{Planar Jet Stripping of Liquid Coatings: Numerical Studies}

\author[ida]{Wojciech Aniszewski\corref{cor1}}
\author[ida]{St\'ephane Zaleski}
\author[ida]{St\'ephane Popinet}
\author[ida,utwente]{Youssef Saade}

\ead{aniszewski@dalembert.upmc.fr,aniszewski@protonmail.ch}

\address[ida]{Institut Jean Le Rond d'Alembert -- CNRS UMR 7190 -- Sorbonne Universite, Paris, France}
\address[utwente]{Physics of Fluids, Faculty of Science and Technology, University of Twente, The Netherlands}

\cortext[cor1]{Corresponding Author at:  Institut Jean Le Rond d'Alembert, Sorbonne Universite,  BC 162, 4 Place Jussieu, 75252 Paris, France. tel: (+33)144-278-714, email: aniszewski@dalembert.upmc.fr, aniszewski@protonmail.ch}

\journal{International Journal of Multiphase Flow}

\renewcommand\Re{{\rm Re}\,} 
\newcommand\We{{\rm We}\,} 
\newcommand\Ca{{\rm Ca}\,} 

\newcommand\be{\begin{equation}}
\newcommand\nd{\end{equation}}
\newcommand\bed{\begin{displaymath}}
\newcommand\ndd{\end{displaymath}}

\newcommand\ba{\begin{array}}
\newcommand\ea{\end{array}}

\DeclareMathOperator{\sgn}{sgn}
\DeclareMathOperator{\erf}{erf}
\definecolor{mygreen}{rgb}{0.9,0.2,0.9}

\begin{document}

\begin{abstract}

  In this paper, we present a detailed example of numerical  study of film formation in the context of metal coating. Subsequently we simulate wiping of the film by a planar jet. The simulations have been performed using \textit{Basilisk}, a grid-adapting, strongly optimized code. Mesh adaptation allows for arbitrary precision in relevant regions such as the contact line or the liquid-air impact zone, while coarse grid is applied elsewhere. This, as the results indicate, is the only realistic approach for a numerical method to cover the wide range of necessary scales from the predicted film thickness (hundreds of microns) to the domain size (meters). The results suggest assumptions of laminar flow inside the film are not justified for heavy coats (liquid zinc). As for the wiping, our simulations supply a great amount of instantaneous results concerning initial film atomization as well as film thickness.

\end{abstract}


\maketitle

\section{Introduction}

\subsection{Jet Stripping of Liquid Coatings}

We present here a numerical  study of the liquid metal coating process. First, liquid film formation on a vertically climbing wall is simulated. Subsequently -- in most cases in the same simulation -- we simulate wiping of the created film by a planar air jet. These processes are of major industrial significance e.g. in metallurgy \citep{takeishi}, photography, painting  and manufacturing of materials  \citep{bajpai}, where the need arises to control the thickness of the deposit. One of the means to establish this control  is the use of a airflow, for example with flat planar jets known as ``air-knives''. These, employed above the coat reservoir, will act by thinning the film deposed onto the product in a controlled manner. However, their effect is far less predictable once airflow issuing from them becomes turbulent, especially around product edges. Similarly, significant kinetic energy of the incoming turbulent airflow may cause unwanted coat atomization, forcing the operators to lower injected air velocity below certain thresholds which are in practice found empirically.  Thus, opportunity arises to optimize the industrial process -- at the very least, there is a sustained need for studies of such a configuration.

The process of film formation, which is the basis of the coat formation procedure, has been studied both experimentally and analytically by many authors starting with now classical results of \citep{llfilm}. Analytic solutions were found e.g. by \citep{groenveld70} who focused on withdrawal with ``appreciable'' inertial forces (relatively high \Re flows) or \citep{spiers73} who has modified the withdrawal theory of Landau and Levich, obtaining improved predictions for film thickness that were also confirmed experimentally.   Later,  \citep{snoe} investigated extensively the film formation regimes in which bulges are formed, focusing on the transition between zero-flux and LL-type films.

In the process of coating, liquid is drawn from a reservoir onto a retracting sheet, forming a coat characterized by phenomena such as longitudinal thickness variation (in 3D) or waves akin to that predicted by Kapitza \& Kapitza  \citep{chang94} (visible in two dimensions as well). While the industry standard configuration for Zinc coating is marked by coexistence of medium Capillary number (Ca=0.03)  and film Reynolds number $\Re_f >  2000$, we present also parametric studies in order to establish if our numerical method  influences the film regimes obtained in the target configuration. Note that metallurgical effects (solidification) are neglected, as they don't play a role in the initial stages of film formation \citep{hocking}.

Once a stable film is formed on the retracted sheet, it can be further thinned  by striping/wiping with airflow. The latter, in most cases, will be a turbulent flow, as the high \Re in the gas are required to exert sufficient pressure on the liquid coat. Although the airflow effects on the coat  can be studied using time averaging \citep{myrillas}, certain instantaneous effects, such as forming of bulges and/or edge effects will not be accounted for. Thus, numerical simulations are a promising tool to supplement experimental studies in this field.   One of the first systematic accounts of the jet stripping of liquid coatings comes from \citep{ellenvu} who have shown analytically that not only pressure gradient acting on the film, but also surface shear stress term plays an important role in the coat thickness modification. \citep{tuck} derived analytical expressions for a dependency between jet airflow velocity and resulting film thickness -- assuming only pressure gradients plays role in film deformation --  and adopting the lubrication approximation for the film flow. The work \citep{takeishi} provided numerical solutions for velocity and shear profiles at the film-air interface  during wiping (using a glycerine solution as the coating liquid).

In 2017 the authors of \citep{hocking} has analysed the problem numerically using a simplified model (including empirically determined shape functions) and a method of lines to study the modified equations of \citep{tuck}. They concluded, for example, that disturbances   to the coating (including bulges and dimples) above the impact zone will persist more likely for thinner coats, as thick ones 'compensate' for that with surface tension and solidification intensity.

In this work, we follow the DNS  \citep{tsz} approach, i.e. we solve a complete set of Navier Stokes equations describing the flow in both phases (in the one-fluid formulation  \citep{delhaye}) with proper boundary conditions, if permitted by the computational code used. A similar approach has previously been adapted e.g. by \citep{lac06}, however their 2006 paper was limited to two-dimensional  Large Eddy Simulation approach. Still, they were able to recover pressure profiles of an impinging jet, as well as give some rudimentary prediction of the splashing which takes place below the impingement area. The authors of \citep{myrillas} performed a study  very similar to \cite{lac06} -- but using parameters of dipropylene glycol as a coating liquid -- yielding e.g. profiles of the film in the impingement zone. An even more basic 2D study using the VOF method  was published in \citep{yu14}, yielding information e.g. about droplet trajectories after impact.  In this paper, we continue such a numerical approach, this time applying a three-dimensional code with very high spatio-temporal resolutions and grid adaptivity.

\subsection{Problem Specification}
\begin{figure*}[ht]
  \centering
  \includegraphics[width=.84\textwidth]{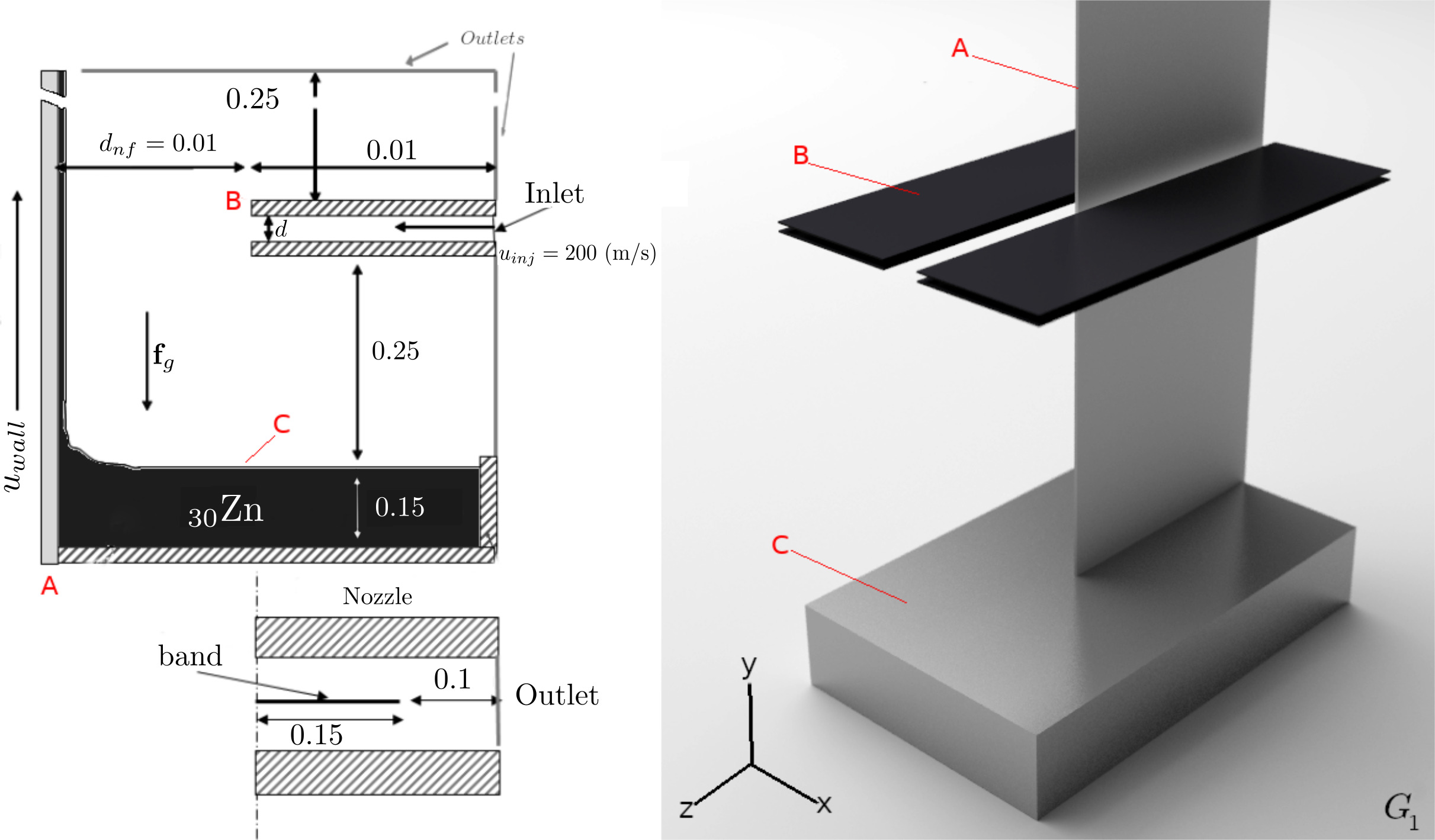}
  \caption{The coating configuration in two (left, half of the geometry visible) and three (right) dimensions. A - upward moving band; B - the ``air-knives'' or flat jet nozzles; C - liquid zinc containers. Note that outer domain walls are invisible in 3D rendering. }\label{2d3dsetup}
\end{figure*}

The investigated configuration is visible in Figure \ref{2d3dsetup}. As we can see in the side-view, the nozzle-band distance $d_{nf}$ is measured at 10mm in industrial configuration. Nozzle diameter $d$ is $1$mm. The proportions in the two-dimensional schematic are forgone for presentation purposes, hence the vertical character is slightly more visible in the 3D rendering. Liquid is drawn from the reservoir C at the bottom, which coats the moving band A. Subsequently, air injected from the nozzle(s) B collides with the coated band A and  leaves the flow domain $\Omega$ below and above the nozzle(s); outlets are  drawn in Figure \ref{2d3dsetup} (left) with grayed lines.

Gravity is taken into account, and upward band velocity is in most cases taken at $u=2$m/s. Regardless of the choice of liquid contained in the reservoir C, band A will be coated, although characteristics of the resulting films will depend strongly on the liquid characteristics. Except where noted,  we have decided to choose liquid zinc as the coating liquids. Properties of $\mbox{ }_{30}$Zn are assumed, that is  surface tension $\sigma=0.7 \lb N/m\rb$, density $\rho_l=6500 \lb kg/m^3\rb$ and viscosity $\mu_l=3.17\cdot 10^{-3} \lb Pa\cdot s\rb.$ Properties of the surrounding gas - which in all cases is air -  are density $\rho_a\approx 1.22 \lb kg/m^3\rb$ and viscosity $\mu_a=2.1\cdot 10^{-5} \lb Pa\cdot s\rb.$

As explained below, we introduce multiple sets of boundary conditions in three dimensions. To concisely refer to them, we introduce the following nomenclature to designate the investigated configurations. Three major geometries considered will be termed $G_i$ with $i=1,2,3.$

If present, the second lower index may be used to designate the grid resolutions used. This index will equal the power of two corresponding to the maximum refinement used by the \textit{Basilisk} code described further. And so, for example, $G_{1,14}$ stands for the first configuration at $2^{14}$-equivalent  refinement level.  Most of the distinguishing features of the three geometries have been delineated in Table \ref{Gtab}. In case other quantities (such as injection velocity $u_{inj}$) are varied between configurations, it will be designated in parenthesis (example: $G_{3,11}(u_{inj}=42.$) Using above terminology, we can now revisit Figure \ref{2d3dsetup}: the configuration presented on the left-hand-side is recognized as $G_2$ in 2D, while the r-h-s of Fig. \ref{2d3dsetup} depicts $G_1.$ 

\begin{table*}
  \begin{center}    
    \begin{tabular}{c c c c c c c c} 
      Conf.   & $L_x\times L_y\times L_z$    & $h_w$          & $x_{wall}$  & $u_{wall}$ & $\mathbf{f}_g$ & $\#$ nozzles & $d$ \\ \hline
      $G_1$   & $0.25\times 0.65\times 0.25$ & $1\cdot 10^{-3}$    & $0.125$    & $2$    & $9.81$ & $2$ & $1\cdot 10^{-3}$       \\
      $G_2$   & $0.512^3$                     & $0.5\cdot 10^{-3}$  & $0$        & $2$    & $9.81$ & $1$ & $1\cdot 10^{-3}$       \\
      $G_3$   & $0.0512^3$                   & $0$                 & $0$        & $0$    & $0$    & $1$ & $1\cdot 10^{-3}$    \\
    \end{tabular}
    \caption{Distinguishing features of the $G_1$, $G_2$ and $G_3$ initial conditions (in all dimensions in meters)).}
    \label{Gtab}
  \end{center}
\end{table*}

Our departure point is the full ''industrial'' configuration $G_1$, visible in Fig. \ref{2d3dsetup} on the right. As sketched in Figure \ref{2d3dsetup}, we orient the geometry so that $y$ is the vertical direction, and air injection takes place along $x$ axis with nozzles extended in the $z$ directions. As visible in Table \ref{Gtab} this configuration involves both ''air-knife'' nozzles; additionally there are outlet areas at the $z+,$ $z-$ and $y+$ domain walls. Split boundary conditions are used to ensure that fluid outflow takes place e.g. only above liquid bath level. As shown in the table, the thickness $h_w$ of the coated band is kept at $0.001$m, and the $x-$centered wall moves up $u_{wall}=2$ (m/s). Due to the fact that the $z-$extent (depth) of the coated all is smaller than the nozzle depth, the $G_1$ configuration allows the air issues from both nozzles to collide.

\begin{figure}[ht]
  \centering
  \includegraphics[width=.42\textwidth]{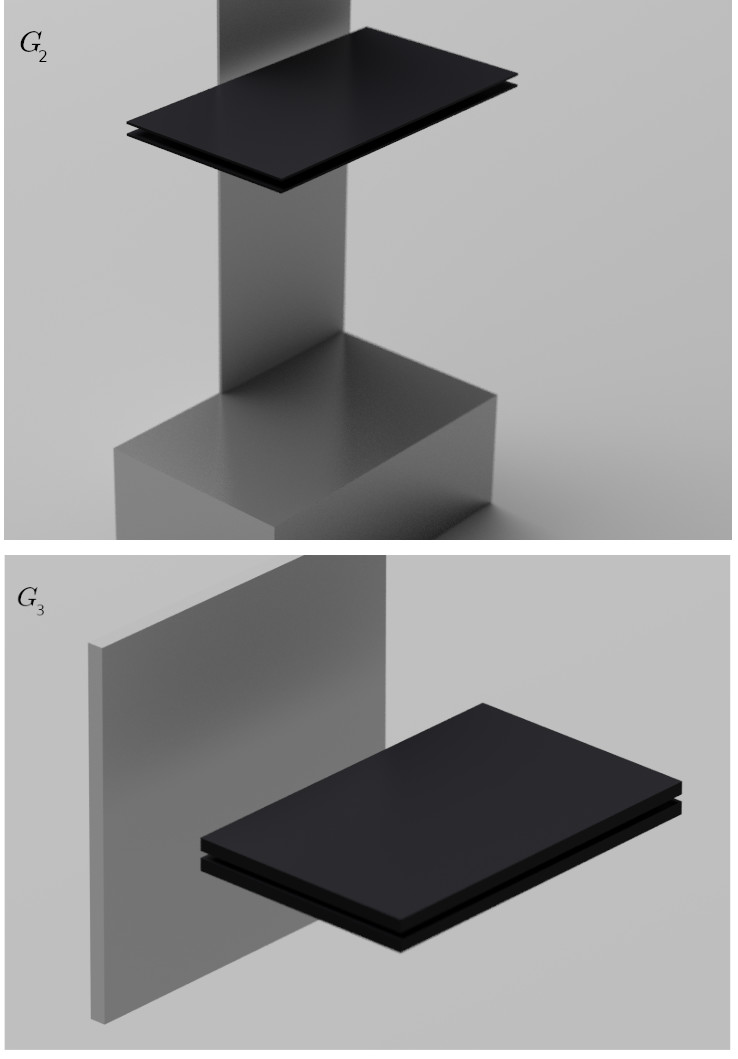}
  \caption{Schematic renderings of the $G_2$ and $G_3$ simulation configurations. Outer domain boundaries are not visible, nozzles are visible in black.}\label{g2g3}
\end{figure}

Two additional configurations are rendered in Figure \ref{g2g3}. As with Figure \ref{2d3dsetup}, note that rendering is not fully up-to-scale: dimensions used in actual simulations are given in Table \ref{Gtab}. The $G_2$ configuration has been created from $G_1$ by including only half of the latter and a symmetry boundary condition at the $x-$ direction. The width ($z-$extent) of the coated wall has also been slightly decreased (from 15 to 5 centimeters) to limit computational cost. Still in the $G_2$ configuration the film is formed gravitationally and the airknife-liquid interaction is maintained. Since the coated wall is placed exactly at $x=0,$ only half of its width is included in the $G_2$ configuration, which makes $G_2$ less suited for studies e.g. of the edge effects of the coated band. Instead, more computational resources can be directed at studying the air-liquid interactions. 

The third introduced configuration, $G_3$ is shown at the bottom of  Figure \ref{g2g3}. It is a ''synthetic'' sub-problem, designed for a fully academic investigation of the air-metal impact phenomenon,  and the initial stages of the two-phase flow post-impact. This is made possible by further reducing the domain size, which now is limited to a $0.05\times 0.05\times 0.05=0.000125\mbox{m}^3$ cube, encompassing a $0.05$m deep ($z$-extent) fragment of the flat nozzle, a nozzle-film gap and the coat. The coated wall is not present except as a boundary condition on the $x-$ direction. In this configuration, film is pre-defined (at thickness $h_{00}$ as explained below) and no gravitational coating is present. A combination of outflow and symmetry boundary conditions are used on all domain walls, with the exception of a partial inlet at the  $x+$. Using the $G_3$ configuration, we further reduce the associated cost of simulating the in-nozzle flow as well as the gas-liquid impact.

\section{Description of the Flow}
\subsection{Governing Equations}

In all cases presented here, full Navier-Stokes equations:

\begin{equation}\label{nswork}
  \frac{\partial {\ub}}{\partial t} + \nabla \cdot \left({\ub}\otimes{\ub}\right)=\frac{1}{{\rho}}\left(\nabla\cdot\left({\mu}{\db}-{p}\ib \right) + {\sigma}\mbox{ }{\nb}\kappa\delta_S \right) +\mathbf{f}_g,
\end{equation}

are solved, assuming also incompressibility of the flow:

\begin{equation}\label{cont}
  \nabla\cdot\ub=0.
\end{equation}

In (\ref{nswork}), $\ub$ stands for the velocity and $p$ signifies pressure. Liquid properties (which vary with the phase) are designated $\mu$ and $\rho$ for viscosity and density, respectively.  Symbols $\ib$ and $\db$ represent unitary and rate of strain tensors, respectively, with $\db$ defined as \[\db=\nabla\ub+\nabla^T\ub.\] Body force (gravity) is taken into account and represented by $\mathbf{f}_g.$ We will occasionally refer to the directions ``up'' and ``down'' which in both two- and three-dimensional simulations are to be associated with $y$ axis. Surface tension is taken into account into the presented simulations, and represented in (\ref{nswork}) by $\sigma\nb\kappa\delta_s$ where $\sigma$ is a coefficient, $\kappa$ is the curvature of the interface $S$, while $\delta_S$ is Dirac distribution centered on it. We assume a one-fluid approach \citep{delhaye}, in which density and viscosity can change at $S,$ and a pressure jump is possible there in case of nonzero surface tension. Below, we will occasionally denote fluid properties with suffixes $l$ and $g$ (liquid/gas) to denote phases.

\subsection{Gravitational Film Formation}

Regarding the film formation on a moving wall similar to A in Figure \ref{2d3dsetup}, we may assume, after \citep{groenveld70} that for a film with locally constant thickness $h,$ equations (\ref{nswork}) simplify to

\begin{equation}\label{gr1}
  \mu_l \frac{\partial^2 u_y}{\partial x^2} = \rho_l g.
  \end{equation}

Integrating (\ref{gr1}) one obtains a parabolic profile inside the vertically moving film

\begin{equation}\label{gr2}
  u_y=\frac{\rho_l g x^2}{2\mu_l}+\mathscr{C}\frac{x}{\mu_l},
\end{equation}
with arbitrary $\mathscr{C}\in\mathbb{R}.$ Using this profile, one can define a dimensionless flux

\begin{equation}\label{gr3}
  Q\colon= \frac{q}{u_y}\sqrt{\frac{\rho_l g}{\mu_l u_y}}
\end{equation}

and dimensionless film thickness

\begin{equation}\label{gr4}
  T\colon= h \sqrt{\frac{\rho_l g}{\mu_l u_y}}
\end{equation}

subsequently establishing a following dependency between the two:

\begin{equation}\label{gr5}
Q = T \left( 1-\frac{1}{3}T^2\right).
\end{equation}

Note that knowing the upward-moving wall velocity $u_{wall}$ and liquid properties, finding $h$ is possible from (\ref{gr4}) given that $T$ has been pre-computed and  the flow regime, governed mainly by film Reynolds number $\Re_f$, is applicable. We will employ this to estimate the Groenveld's thickness ($h_G$) of the film when studying its formation in Section \ref{2dffchapt}.

\subsection{Liquid-air interaction}\label{nthsect}

Once a stable film is formed on the substrate, it is acted upon by airknives, a process we will briefly discuss below. In general terms (especially for situations when velocity profile inside the film can be assumed known), the approach to modeling air-liquid interaction is to write the film equation for $h=h(y)$. This equation needs to include terms representing gravity, surface tension, as well as pressure gradient $\frac{\partial p}{\partial y}$and the shear stress $\tau_{yy}.$ Comparing magnitudes of pressure and shear stress with that of surface tension often results in dropping the latter from the model. Subsequently, film equations are solved (steady state solutions are sought) and thickness predictions given, dependant upon $\partial_yp$ and $\tau_{yy}.$ The problem is formulated in such a way that the boundary condition imposed on the moving wall ($x=0$) is $u_y=u_{wall}$. At the interface $(x=h)$, author \citep{hocking} expect \[u_y=\tau_{yy} \frac{h_{00}d}{\mu_l d}\] as a velocity condition\footnote{At this stage, $h_{00}$ can be seen as a order-correct prediction of film thickness, Hocking uses a value of one micrometer.}, and \[p-p_{air}(y)=\sigma\kappa=\sigma \partial^2_{yy}h\] for the pressure. This is supplemented by a transport condition for thickness \begin{equation}\label{nth1}\partial_t h+u_y \partial_x h = u_y.\end{equation}

With these  assumptions, Hocking et al. \cite{hocking} use  linearization and a thin film assumption (treating film thickness $\epsilon$ as an infinitesimal) to simplify the governing Navier-Stokes equations to:

\begin{equation}\label{nth2}
  \partial_t h + \partial_y\left( h+\frac{h^2 G(y)}{2} -\frac{1}{3}h^3\left(S+\partial_yp(y)-C\cdot h_{yyy}\right)\right)=0.
\end{equation}

In the above, $S$ is the Stokes number -- in the cited work, \[S=\frac{\epsilon^2 \rho_l g h_{00}^2}{\mu u_{wall}}\] and is of minus-fourth order --  while $G(y)$ and $P(y)$ are dimensionless shear stress and pressure distribution functions. For this model, interesting empirically established forms of $G$ and $P$  are presented by \citep{tuu} for pressure:

\begin{equation}
  \label{nth3}
  P(y) = P_{MAX}\left(1+0.6y^4\right)^{-3/2},
\end{equation}

and by \citep{elsaadawy} for the shear stress:

\begin{equation}
\label{nth4}
G(y) = \begin{cases} \sgn(y)G_{max}\left(\erf(0.41|y|)+0.54|y|e^{0.32|y|^3} \right) &\Leftrightarrow |y|<d^* \\
                     \sgn(y)G_{max}\left( 1.115 -\log|y|\right) &\Leftrightarrow |y|>d^*  \end{cases} 
\end{equation}

with $d^*\approx 1.73d.$ Naturally, $d^*$ should approximately correspond to the height  of air impact zone, while $P_{MAX}$ and $G_{MAX}$ should be calibrated by supplying correct values associated with gas velocity. (One observes distribution (\ref{nth3}) to be similar to simulated pressure bell curves e.g. in Fig. \ref{synt_prof_p}a). As far as (\ref{nth2}) is considered, \citep{hocking} apply $\partial_t h=0$ and $h_{yyy}=0$ as discussed above. Note that the expression in parenthesis in (\ref{nth2}) is the flux $q$ of coating material; within a thin-film approximation we might request $\partial_h q=0$ at certain critical points, as well as $\partial_y q=0$ thus finding $h.$

A slightly different way of obtaining the $\partial_h q =0$ constraint is to first assume a known, e.g. Poiseuille velocity profile inside the film. This could be written as

\begin{equation}
  \label{nth5}
  u_y(x,y)=u_{wall}-\frac{\rho_l g+\partial_y p}{2\mu_l}x(2h-x)+\frac{\tau_{yy}(y)x}{\mu_l}.
\end{equation}

We integrate (\ref{nth5}) to get the flux equation:

\begin{equation}
  \label{nth6}
  q = u_{wall}\cdot h - \frac{h^3\left(\rho_lg +\partial_yp\right)}{3\mu_l}-\frac{\tau_{yy}h^2}{2\mu_l}.
\end{equation}

Note that in (\ref{nth6}) the zero-flux thickness $h_{00}$ can be obtained by zeroing pressure gradient and shear strain terms. Similarly to \citep{hocking} we can now request the flux (\ref{nth6}) to have zero derivative with respect to $h$ leading to \[ \mu_l u_{wall}-A_1h^2 - \tau_{yy}(y) h =0,  \] yielding a solution for thickness we denote $h_c\colon$

\begin{equation}
  \label{nth7}
  h_c=\frac{\mu_l u_{wall}}{\tau_{yy}(y)}\left(1+\frac{A_1 u_{wall}\mu_l}{\tau^2_{yy}(y)}\right),
\end{equation}

where $A_1 = \rho_l g + \partial_y p.$ We rephrase the above result by introducing a pressure to shear strain ratio $\epsilon_3$ defined as \[ \epsilon_3 = \frac{A_1 u_{wall} \mu_l}{\tau^2_{yy}(y)};\] with this, (\ref{nth7}) becomes

\begin{equation}
  \label{nth8}
  h_c=\frac{\mu_l u_{wall}}{\tau_{yy}(y)}\left(1+\epsilon_3\right). 
\end{equation}

Further on, by approximating pressure and strain, for example by  \begin{equation}\label{nth8b}p_{air} = c_p \rho_g u_{inj}^2\end{equation} we can represent $\epsilon_3$ as \[ \epsilon_3 = \frac{c_p}{c_s}\cdot\frac{u_{wall}\mu_l/\mu_g}{\Re_g\mu_g},\]

with $c_p$ and $c_s$ being pressure- and shear strain-related dimensionless constants. Even if we -- somewhat optimistically -- calculated $\Re_g$ using $d$ as reference length, in the discussed applications we still could have $Re_g \gg 10 \Rightarrow \epsilon_3 \ll 1,$ which would reduce (\ref{nth8}) to \[ h_c \simeq \frac{\mu_l u_{wall}}{\tau_{yy}}.\]
This simple result relates thin-layer-approximated thickness to shear-stress; the actual  thickness approximations for $u_{inj}\approx 200$  using that formulation will be given below.  In the general, non-laminar case,  the relation between $q$ and $h$ cannot be established beforehand. Still, as a working measure, we can define the average velocity $\bar{u}$ in the reference frame of the moving wall, such that \[ q = (u_{wall}-\bar{u})h.\] Using $\bar{u}$ one may examine the balance of air and wall stresses, pressure and shear strain, viscosity and gravity in the form

\begin{equation}
  \label{nth9}
  \rho_l g h+c_p\rho_gu_{inj}^2h/d = c_f\rho_l\bar{u}^2+\frac{3\mu_l\bar{u}}{h}-\frac{3\tau_{yy}}{2},
\end{equation}

with $c_f$ being a wall friction  dimensionless coefficient. By once again zeroing the flux derivative one obtains a result similar to (\ref{nth8}) with a $c_f$-related correction

\begin{equation}
\label{nth10}
h_c=h_{c,laminar}(1+\epsilon_5)
\end{equation}

where
\begin{equation}
  \label{nth11}
  \epsilon_5 = \frac{c_f}{c_s}\frac{(\rho_l/\rho_g)u_{wall}^2}{u_{inj}^2}.
  \end{equation}

While it is reasonable to expect $c_f<c_s$ the actual estimates are nontrivial to obtain; we can however conclude that films thicker than in the laminar case  are possible in this regime. Nevertheless, post-impact film thickness $h_c$ values of order of microns should be expected, which pose a significant challenges to computational simulations. Additionally, it must be noted unsteady solutions, as presented in two dimensions by \citep{hocking} involve wavy structures pushed  away from the impact zone; while we don't present a quantitative description of such dips and depressions, their appearance is expected. This further complicates the task of establishing effective coating thickness $h_c$ above the impact zone. Our estimates of $h_c$  will be given below (see Section \ref{YSsect}), as well as summarized for the industrial parameters in Table \ref{thicktab}.

\section{Computational methods}\label{methodsect}

In the research presented here we have applied the ``Basilisk'' computational code \citep{sp15}, which is an in-house, GPL-licensed code whose main developer is one of the present authors (SP). It is a descendant of the ``Gerris'' code  \citep{popinet2} and as the latter, it enables local adaptive mesh refinement (AMR)  \citep{pill32} using and quad/oct-tree type mesh (regular, structured cubic meshes without refinement are also possible). The code is optimised for speed (which differs it from Gerris) and capable of both OpenMP (single node) parallelism and MPI-type (multi-node) operation. Most recently, Basilisk has been applied e.g. to model compressible flows connected to bubble dynamics \citep{fuster2018}, propose single-column models in meteorological simulations \citep{antoon2018}, or simulate turbidity currents \citep{yang18}.  We conclude our description of the code by briefly remarking about two features that make it stand out: firstly, it is a multi-equation solver, i.e. a broad framework that allows choosing equations to be solved, making it de-facto a multi-physics code. Secondly, it contains a built-in parser/lexer providing ``targeted'', minimal re-compilations for the configuration currently used.

Navier-Stokes equations are solved using a well known projection method  \citep{tsz} with a procedure similar to that applied in Gerris  \citep{popinet1, popinet2}. Centered discretization is used for all scalar and vector fields, with additional face-centered values defined for $\ub$ which are used e.g. to ensure divergence-free condition during mesh refinement. For consistency reasons, advection term of (\ref{nswork}) is defined and calculated on cell faces as is $\nabla p.$ Advective fluxes are obtained using the Bell-Collela-Glaz advection scheme  \citep{bell89}. Discretizations are generally finite-differencing up to second order unless noted otherwise. The Runge-Kutta scheme is used for time advancement, and a certain optimisation of Poisson equation's solution  is given by implementing the Multigrid (MG) method  \citep{brandt}.

The Volume of Fluid (VOF) method  \citep{tsz} is used to track the interface using geometric interface reconstruction  \citep{aniszewski2014caf}. In this method, fraction function $C$ (equal to one or zero in either phases ) is passively advected with the flow. Grid cells with fractional $C$ values are those in which interface is geometrically reconstructed and represented by a line/plane (in two and three dimensions, respectively). Note that $\mu$ and $\rho$ are usually local functions of $C.$ Interface curvature is also computed from $C,$ using the Height-Functions method  \citep{afkhami08, popinet2} taking into account proper treatment close to solution boundaries.

Basilisk's procedure for local mesh adaptation employs a wavelet transform of a given scalar field to assess the latter's discretization error. If the error is above the user-specified threshold, the grid is locally refined by subdividing it onto four (quad-tree) and eight (octree) sub-cells and performing a \textit{prolongation} of the courser-mesh scalar onto children cells to obtain their initial values (the inverse process is termed \textit{restriction}). For the simulations presented herein, we use $\ub$ and $C$ fields' error as the refinement criteria with $1\cdot10^{-3}$ and $1\cdot10^{-2}$ error thresholds, respectively.

\subsection{Ensuring Momentum Conservation in Two-Phase Flow}\label{momconssect}

The momentum-conserving methods  \citep{geo2015}  derive from a variant of VOF  \citep{hn} method originally proposed in   \citep{rudman} to treat two-phase flows with considerable density ratios. General idea is that instead of a simple incompressibility assumption

\begin{equation}
  \label{incomp}
  \nabla\cdot\ub=0,
\end{equation}

we instead write the mass transport equation in full, as is done in compressible formulation  \citep{pilliod_phd}:

\begin{equation}
  \label{masst}
  \frac{\partial \rho}{\partial t} + \nabla\cdot (\rho \ub) = 0,
\end{equation}

using also the conservative form of the Navier-Stokes equations (not shown) \citep{geo2015}, which contain the momentum term \begin{equation}\label{moment}\nabla\cdot(\rho\ub\otimes\ub).\end{equation} Subsequently, in implementation, we calculate density \textit{from the fraction function} definition:

\begin{equation}
  \label{densc}
  \rho=\rho_l C + (1-C)\rho_g,
\end{equation}

instead of the other way around as it is done traditionally  \citep{hn}. The way in which the momentum-conserving methods stand out from traditional two-phase Navier-Stokes equation models using VOF is that subsequently, the $\rho(C)U$ products found in both (\ref{masst}) and (\ref{moment}) are calculated consistently in the same control volumes. This can be non-trivial if staggered grid  \citep{tyliszczakJCP} discretizations are used, and can be solved either by grid-cell subdivisions \citep{rudman} or using sub-fluxes of fraction function \citep{geo2015}. Thus consistency between transports of mass and momentum are ensured numerically, resulting in a far more robust computation.

\subsection{Implementation of embedded solids}\label{solidsect}

Problem geometry illustrated in Figure \ref{2d3dsetup} is nontrivial, due to the fact that  flow is expected to take place around walls of the coated band, as well as the edges and corners defining the flat nozzle, i.e. space containing embedded (or immersed) solids, and the computational code used must allow for this. We use a rudimentary technique of locally modifying the velocity field for this purpose. Local modification of scalar fields is a relatively simple technique used when simulating large-scale systems involving solids  \citep{chinesebird}. It is a strongly simplified variant of the Immersed Boundary Method (IBM) of Peskin  \citep{peskin}, which does not modify the grid data structure and is thus compatible with MPI protocol. If we denote the interior of the solid contained by boundary $\Gamma$ by $\overline{\Omega}$ we can note:

\begin{equation}
  \label{chinapower}
  \forall \xb \in \overline{\Omega}\cup\Gamma\colon \ub(\xb)=0,
\end{equation}

that is, all velocity components are  set to zero within the solid. As long as no provisions are needed for $x\in\Gamma,$ the crude approximation provided by (\ref{chinapower}) yields satisfactory results  \citep{chinesebird}. A moving wall can be prescribed by using a non-zero ($u_{wall}$) right-hand side in (\ref{chinapower}). Note however, that pressure $p$ is not modified in any way inside the solid $\overline{\Omega}$ which, in principle, may result in its incorrect values especially at boundary $\Gamma.$ This could be addressed for by locally modifying pressure gradients, which in a physical sense is equivalent to defining a certain force which would only be nonzero at the boundary  \citep{gibouibm}. This however complicates the technique to a degree comparable with implementation of domain reshaping, as optimally, it should be followed by removal of the interior points from the grid.

Instead, we note that for geometries presented -- even the most complicated $G_1$ setup -- the domain interior is merely a sum of cuboids: it contains no inclined nor curved surfaces. The no-slip condition at the surface of the substrate wall moving with velocity $u_{wall}$ can be reasonably approximated using (\ref{chinapower}). Thus, for the current calculations we adopt this simple technique. 

\subsection{Spatially Restricted Refinement}

To limit the associated CPU cost of the grid refinement, we have employed additional technique of spatially restricted refinement  (for short, we will use the abbreviation 'SRR' below). Using SRR is straightforward. The quad/oct-tree data structure in \textit{Basilisk} results in subdivisions of cells into four/eight sub-cells as the grid is refined in two or three dimensions respectively. The entire domain is a 0-level (parent/root) cell with four/eight 1-level sub-cells and so on. The subdivision is performed based on criteria stemming from error estimation on chosen scalar fields performed in wavelet space (usually, components of $\ub$ and/or VOF fraction function). If refinement criterion is locally fulfilled, \textit{Basilisk} will keep refining the grid up until reaching the maximum allowed level. The SRR technique locally limits this maximum grid level using a spatial criterion. This means larger discretization errors are intentionally allowed far from regions of interest. Thę latter regions have to be predefined before the simulation. Then, dynamic grid refinement will act as usual, the only difference being that refinement to maximum level will take place only in chosen domain sub-areas while outside of them lower maximum level is forced. This tactic of refinement situates the presented simulation between the block-based \citep{lakehal2010} and point-based \citep{popinet2} mesh refinement.

\section{Results}
\subsection{Planning of the Simulations}

The full air-knife configuration poses numerous challenges for computational simulation. Firstly, it comprises of sub-processes -- such as interaction of  turbulent structures with planar interface -- which are very demanding on their own. This is either for reasons pertaining code stability \citep{geo2015}, reliability of results  \citep{sailor_jet} or CPU-cost \citep{aniszewski}. Secondly, the geometry of the problem, as presented in Figures \ref{2d3dsetup} and \ref{g2g3} results in a complicated flow. The latter includes a range of scales -- from domain size to liquid sheet thickness -- that require very fine resolution. However resolution could be limited  only in region of interest, which amounts to a relatively small part of the simulation domain. For this reason, we have implemented a broad campaign of simulations focused on individual sub-problems. For reasons of brevity we will only present here a subset of the obtained results, namely:

\begin{enumerate}
\item A film formation study: $G_{1,11}$ (w/o the air injection nozzles) and $G_{2,14}$ in 2D;
\item A brief, 2D validation on the dynamics of the jet impinging on flat plate ($G_{2,14}$ w/o the liquid phase);
\item Studies of film formation and airknife-liquid interaction with ''relaxed'' and industrial parameters.
\item Simulations of the full configuration using $G_2$ and $G_3$ geometries with varying spatial resolutions and injection velocities.
\end{enumerate}

In the above the ``relaxed''  parameters simulations assume a decreased $\We$ and $\Re$ as a means of preparation, converging to the final solution with increasing dimensionless numbers. For reference, Table \ref{realrelax} contains parameters for both industrial and relaxed parametrisations of the considered problem. Most important differences include an order of magnitude lower liquid density and higher $u_{wall}$: both of these contribute to sway the balance between gravity and liquid uptake towards the latter. This subsequently leads to a thicker film formed, thus decreasing associated CPU cost needed to perform simulation. (For the same reasons, in gas phase, velocity  $u_{inj}$ is decreased twofold in relaxed parametrisation.) This results for example in the zero-flux $h_{00}$ thickness of the film in relaxed parameters being fourteen times that of its value in industrial parameters.

\begin{table*}[h!]
  \centering
  \begin{small}
    \begin{tabular}{ l  c c l c c c c c  }
      \hline
      Case    &    $\rho_l$ & $\rho_g$  &  $\mu_l$      & $\mu_g$       & $u_w$  & $d$ & $d_{nf}$  & $u_{inj}$  \\
      \hline
      Unit    & (kg/m$^3$)  & (kg/m$^3$) & (Pa$\cdot$s) & (Pa$\cdot$s)  & (m/s)  & (m) &  (m)      &  (m/s)     \\
      \hline
      Relaxed & $650$       & $1.22$     & $3.17\cdot 10^{-2}$ & $1.7\cdot 10^{-5}$  & $4$    & $0.001$ & $0.01$   & $75$ \\
      Industrial    & $6500$ & $1.22$  & $3.17\cdot 10^{-3}$ & $1.7\cdot 10^{-5}$  & $2$    & $0.001$  & $0.01$   & $200$ \\
      \hline
    \end{tabular}
    \caption{Parameters for the discussed simulations in industrial  and ''relaxed'' variants).}
    \label{realrelax}
  \end{small}
\end{table*}

Additional difference between the relaxed and industrial configurations is the coated plate thickness, it is held at $5$mm for the relaxed variant and $1$mm in industrial. Nozzle wall thickness is configured analogically. Both changes facilitate the implementation of simulation geometry in the relaxed case, meaning that coarser grids suffice to implement (\ref{chinapower}) formulation as more grid-points end up contained in the $\overline{\Omega}$ region.

\subsection{Film formation studies}\label{2dffchapt}

Film formation studies focus on the steel band emerging from the zinc reservoir. As said above, this is implemented using the domain reshaping technique of \textit{Basilisk}. Such studies allow for observation of e.g. edge effects at the stage well before the initial coat is modified with air-knives  \citep{ellenvu}. Even by studying this problem in two dimensions a lot can be learned e.g. about the flow inside the film. We can define the film Reynolds number, describing internal liquid flow as

\begin{equation}\label{Refilm}
  \Re_f(h_{00})=\frac{\rho_l h_{00} u_{w}}{\mu_l},
\end{equation}

where $\rho_l$ is liquid density, $u_w$ is the upward-moving band (wall) velocity, and $h_{00}$ is a zero-flux film thickness  \citep{groenveld70}, i.e. that at which liquid fluxes associated with wall movement (upwards) and gravity (downwards) balance out. Thickness $h_{00}$ can be found by assuming parabolic velocity profile and comparing dimensionless flux and film thickness, leading to

\begin{equation}\label{h00eq}
  h_{00}=\sqrt{\frac{3\mu_l u_w}{\rho_l g}}.
\end{equation}

Using (\ref{h00eq}) and calculating $\Re_f$ from (\ref{Refilm}) we arrive at values of $h_{00}=5.46\cdot 10^{-4}$ and $\Re=2240$ for industrial parameters. Indeed, one could say simulations prove that industrial-class metal coating is a man made system on the edge of criticality, as this is very close to critical $\Re_f$ values delineating laminar and turbulent film formation regimes. A slightly more delicate interpretation is suggested once we modify our expectations towards film thickness as follows.

Assuming that the withdrawal is dominated by inertial forces, one can employ Groenveld's analysis mentioned in the context of equation (\ref{gr4}). In \citep{groenveld70}, values of $\Re_f$ and $\Ca$ characterizing ''industrial'' parameters place our case in the high-$\Re$ regime, for which Groenveld proposes values of $T$ and $Q$ at $0.52$ and $0.47$, respectively. Using these with (\ref{gr3}) and (\ref{gr4}) one can estimate the associated thickness $h_G=163\mu m.$ We will use this value below as a rough estimate of the expected film thickness for gravitational withdrawal simulated in this work. Using thus calculated thickness value we may modify (\ref{Refilm}) like so:

\begin{equation}\label{RefilmG}
  \Re_f(h_{G})=\frac{\rho_l h_{G} u_{w}}{\mu_l} \approx 672.
\end{equation}

While this value is three times smaller than $\Re(h_{00})$, one could expect the film to be at least in the intermittent regime.

\begin{figure}[ht!]
  \centering
  \includegraphics[width=.5\textwidth]{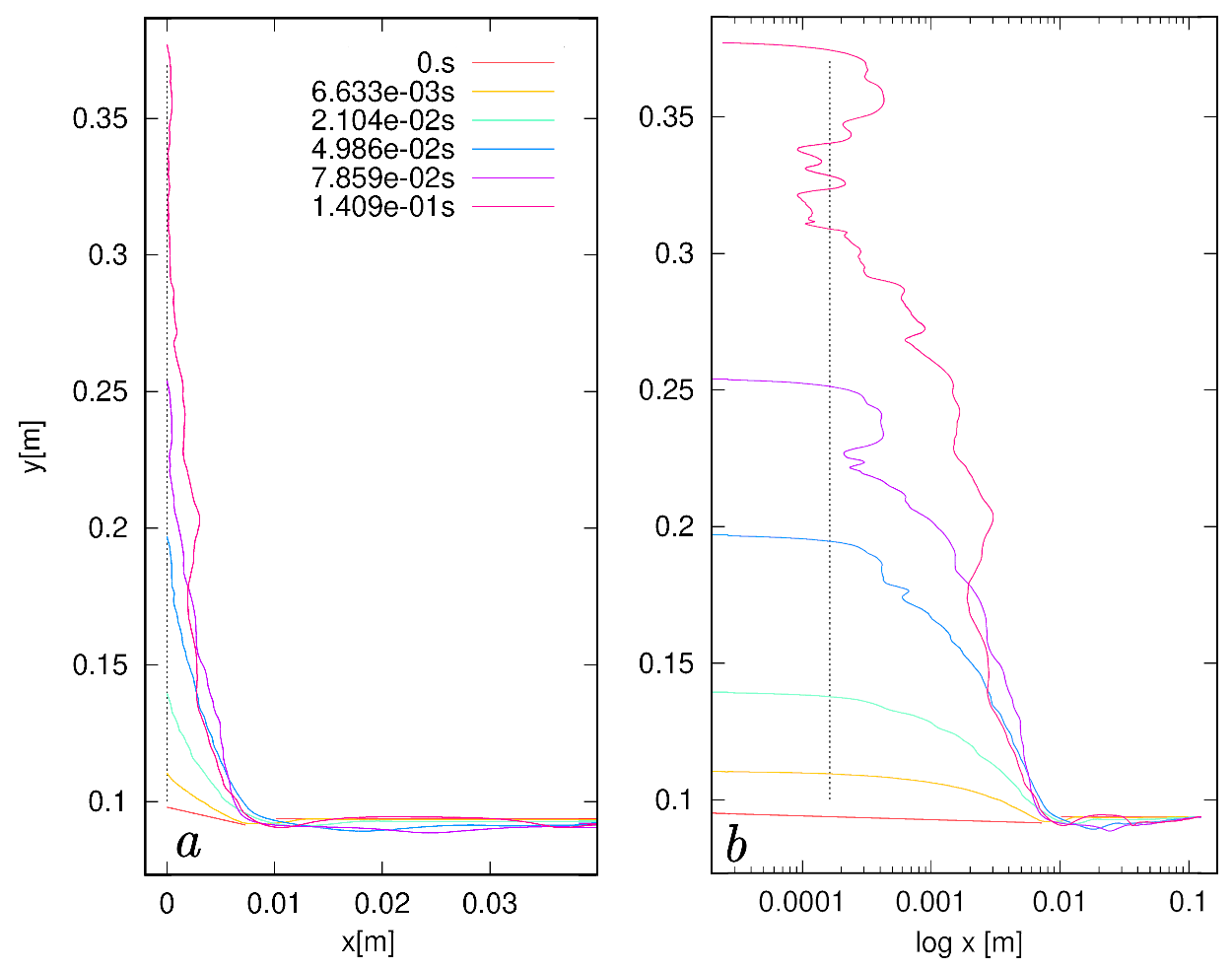}
  \caption{Configuration $G_{2,14}$ (no air injection). Interface geometry at chosen $t$ values (with $x$ in (a) dec and (b) log). The dashed line is $h_G=163 \mu$ m.}\label{2dints}
\end{figure}

Figure \ref{2dints} presents the interface geometry at simulated time values $t<1.409\cdot 10^{-1}$s for the industrial parameters simulation. (This simulation is configured in such way that the upward-moving band is defined as a boundary condition, so no solid embedding technique is needed.) It is visible that in this case the film head penetrated upwards somewhat faster than $u_w$ would suggest; we could attribute it to the boundary condition for the $C$ function  \citep{afkhami18}. Wavy character of the film is easily observable, especially for the final curve corresponding to $t=0.14$s. This is emphasized in Fig. \ref{2dints}b showcasing the very same curves with logarithmic scale used for the $x$ axis. Moreover, dashed line in Fig. \ref{2dints}b, representing Groenveld's prediction using high-$\Re$ theory is reasonably approximated by our result, save for the aforementioned wavy film character. More specifically, the recovered Groenveld's thickness $h_G$ is $163\mu m,$ i.e. less than ten percent of the $h_0$ thickness discussed in  \citep{myrillas} in the context of dipropylene glycol, and requires a substantial grid resolution to resolve. The result visible in Fig. \ref{2dints} has been obtained with $14$ levels of refinement. Domain size has been $L=0.65$m (only a part is visible in Figure \ref{2dints}). Thus, an individual grid  element has the size of \[\Delta=L/2^{14}\approx 39 \mu m,\] resulting in approximately four grid elements per film thickness at its thinnest point.

\begin{figure}[ht!]
  \centering
  \includegraphics[width=.5\textwidth]{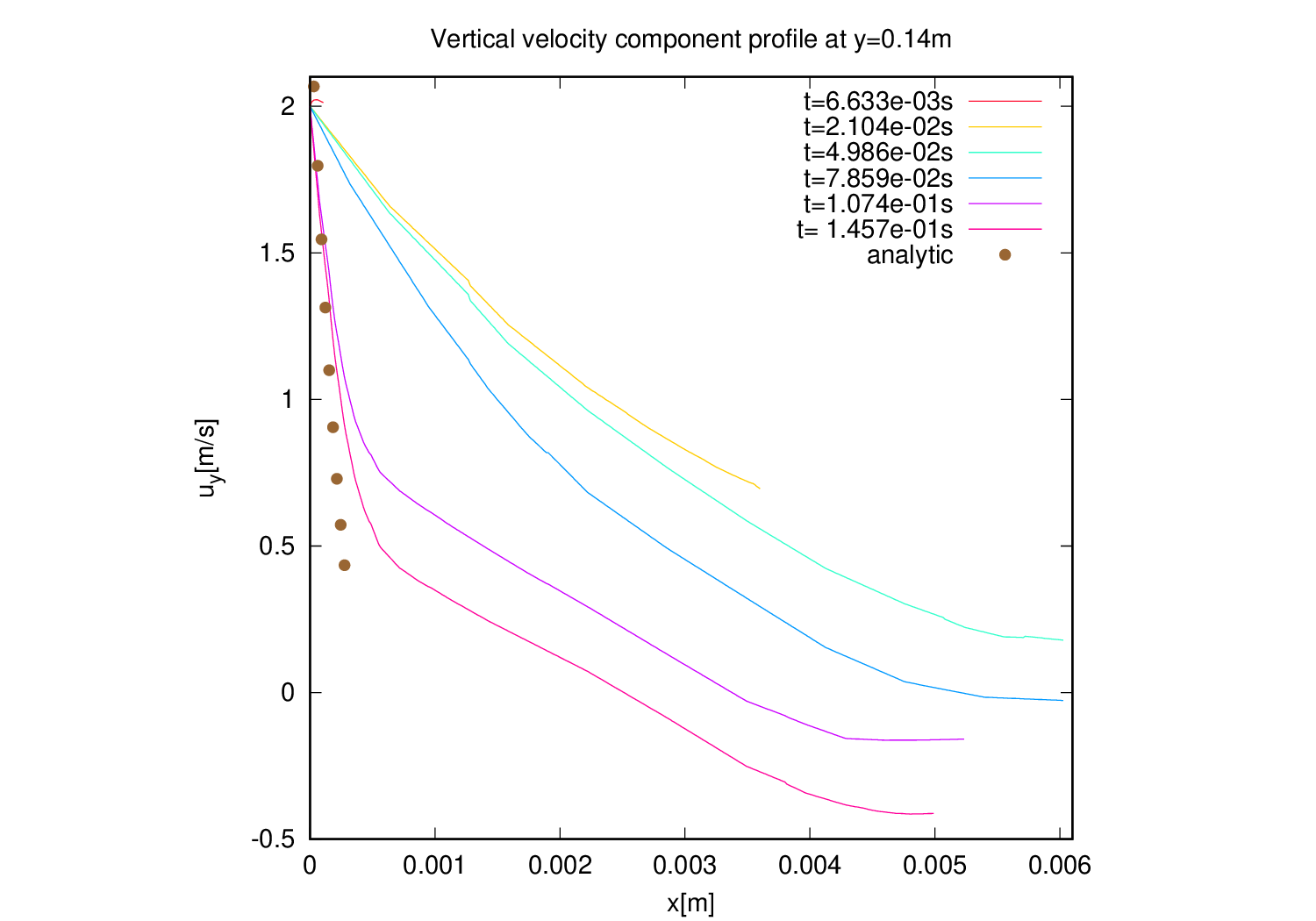}
  \caption{Configuration $G_{2,14}$ (two-dimensional, no air injection). (a) $u_y(x)$ profiles through the film at varying $t$ values taken from Fig. \ref{2dints} (lines), Groenveld's prediction using (\ref{gr2}) (points).}\label{2dprof_thick}
\end{figure}

Figure \ref{2dprof_thick} shows the creation of a boundary layer for the various moments in time (in the $t>0.1457$s range) of the same flow. The profiles have been sampled at  $h=0.14m$ or $0.04m$ above the reservoir.  The velocity profile remains parabolic, however it clearly becomes steeper for $t>0.1$s with an apparent  plateau extending for $x>5\cdot 10^{-4}$ suggesting a detachment of the layer adjacent to the plate  \citep{snoe}. In Fig. \ref{2dprof_thick} we additionally compare the profile for $t=0.1457$s with analytical expression (\ref{gr2}) (dots) using $\mathscr{C}=1.$ Consistency is visible especially closest to the wall, suggesting that the final profiles lend themselves  well to those assumed in \citep{groenveld70}, as hinted previously by Figure \ref{2dints}. This serves as a convincing argument that the film evolution is reasonably well described by the high-$\Re$ theory. %

Moreover, in Fig. \ref{2dprof_thick} profiles are sampled only for $C>0$ (i.e. inside the liquid film). Thus, for each of the lines, the abscissa of its right-hand end-point corresponds to the film thickness $h(y)$ at $y=0.14$m. As one can observe for $t\in \lb 4.9,7.8 \rb$ we have $h(0.14,t)\approx 6\cdot 10^{-3}$m whereas for $t=1.457\cdot 10^{-1}$ the thickness drops, suggesting a bulge has passed over the point and retracted.

\begin{figure}[ht!]
  \centering
  \includegraphics[width=.5\textwidth]{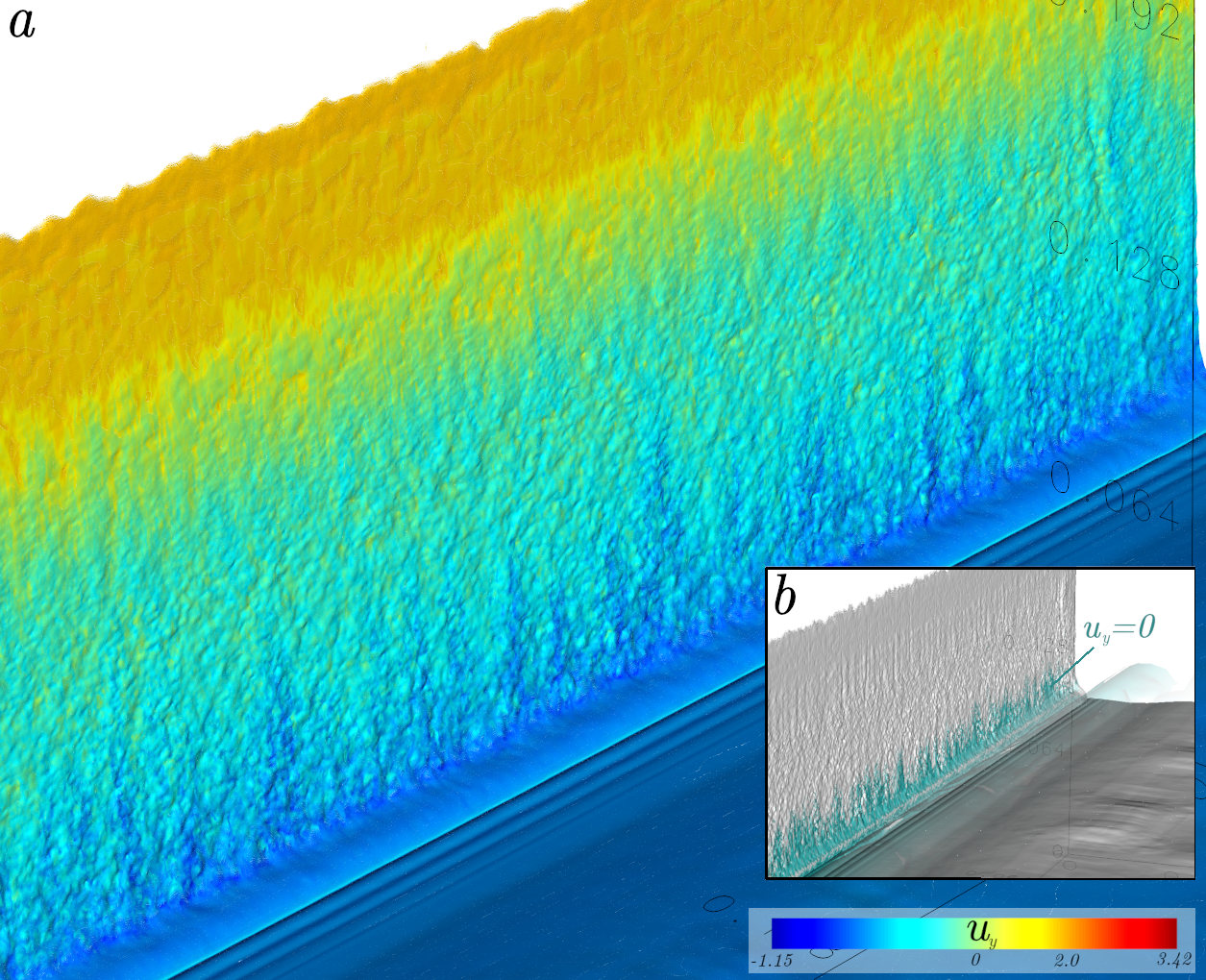}
  \caption{Configuration $G_{2,12}$ film formation study; the flow at $t=0.148$s. (a) Actual VOF-reconstructed liquid-gas interface geometry, colored by the $u_y$ velocity component. Inset (b): liquid-air interface shown in gray with the $u_y=0$ isosurface drawn in turquoise to approximately delimit the stagnation height.}\label{g2ff3d}
\end{figure}

Using the $\left(2^{12}\right)^3$-equivalent grid, we have performed a three-dimensional simulation $G_{2,12}$ to study film formation. Its results are presented in Figure \ref{g2ff3d}, which could be seen as a 3D analog of the interface geometry presented above in Fig. \ref{2dints}a. Similar time instance, $t=1.45\cdot 10^{-1}$s is chosen in Figure \ref{g2ff3d}. A heavily ''rugged'' film surface is easily recognizable in Fig. \ref{g2ff3d}a, in which it has been colored by the vertical velocity component $u_y.$ As we can see, distinct liquid boundary layer develops directly adjacent to the wall, traveling with velocity $u_{wall}$. This is fully consistent with liquid velocity profiles presented in Fig. \ref{2dprof_thick} for $t=1.45\cdot 10^{-1}$s. As we get further from the boundary layer, velocity at which the film is climbing drops sharply; Fig. \ref{g2ff3d}a indicates also that surface material crumbles back into the bath (blue areas close to the reservoir height). We have included, as an inset in Fig. \ref{g2ff3d}b, an isosurface for the zero vertical velocity $\left(u_y=0\right)$, rendered in turquoise against the gray interfacial surface. (Note that $u_y=0$ occurs as well in the gas far from the coated wall. For this reason, parts of the isosurface were removed from Figure \ref{g2ff3d}b artificially to not obscure the view of the coated wall area.) In this way, we are able to approximate the stagnation height for $t=1.48\cdot 10^{-1}$s as $0.13$m e.g. $0.03$m above the bath level. Above this height, all flow is upwards.  The interface formations visible throughout the height of the film surface seem sufficiently resolved and not numerically induced. For example, halfway through the film height in Fig. \ref{g2ff3d} film thickness is of order $0.01$m (or eighty times the grid size at $12$ levels of grid refinement).

\begin{figure}[ht!]
  \centering
  \includegraphics[width=.42\textwidth]{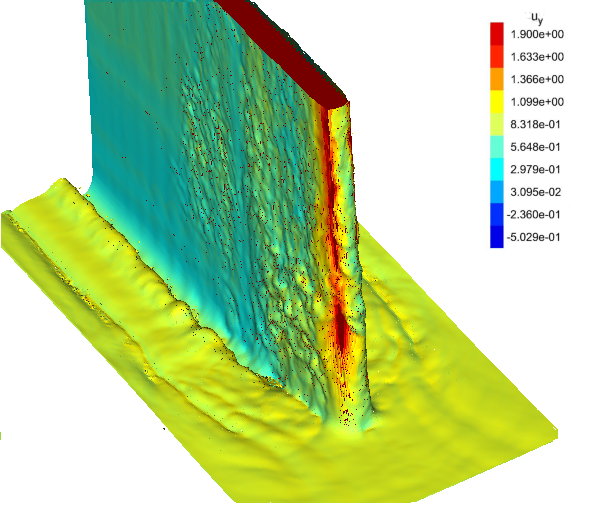}
  \caption{Coating in $G_{1,11}$ configuration (no air injection) at   $t\approx 0.32$s. Interface colored by the vertical velocity component.}\label{evogray2}
\end{figure}

Even using a slightly less refined grid ($11$ levels of refinement, or $2048^3$-equivalent), we still observe a wrinkling of the interface as well, mostly in the $G_1$ configuration which involves the coated edge. The evolution during the phase of fully developed film is visible in Figure \ref{evogray2}, prepared using a $2^{11}$-equivalent grid. In this phase the band is fully coated, while some ``dimples'' appear close to the reservoir surface once zinc is drained. Only a very thin layer of zinc is deposited close to the band edges, as can be seen by the surface color which corresponds to $u_{wall}=2$. The surface of the film undergoes progressive distortion starting from the side of coated band.  This applies especially to the coated $x+$ and $x-$ walls, in which wrinkling appears progressively further from the band edges. The turbulent nature of the film is evident, along with fully three-dimensional character of the wrinkles/waves. To our knowledge, this is the first published result of a 3D coating simulation including the edge, and $\Re_f$ is far higher than the previously published 2D results  \citep{myrillas,lac06}. We observe the wrinkles similarly in the film thickness profiles presented  for all $G_1-G_3$ configurations further on (e.g. Figure \ref{rtis42prof}). 

\begin{figure}[ht!]
  \centering
  \includegraphics[width=.42\textwidth]{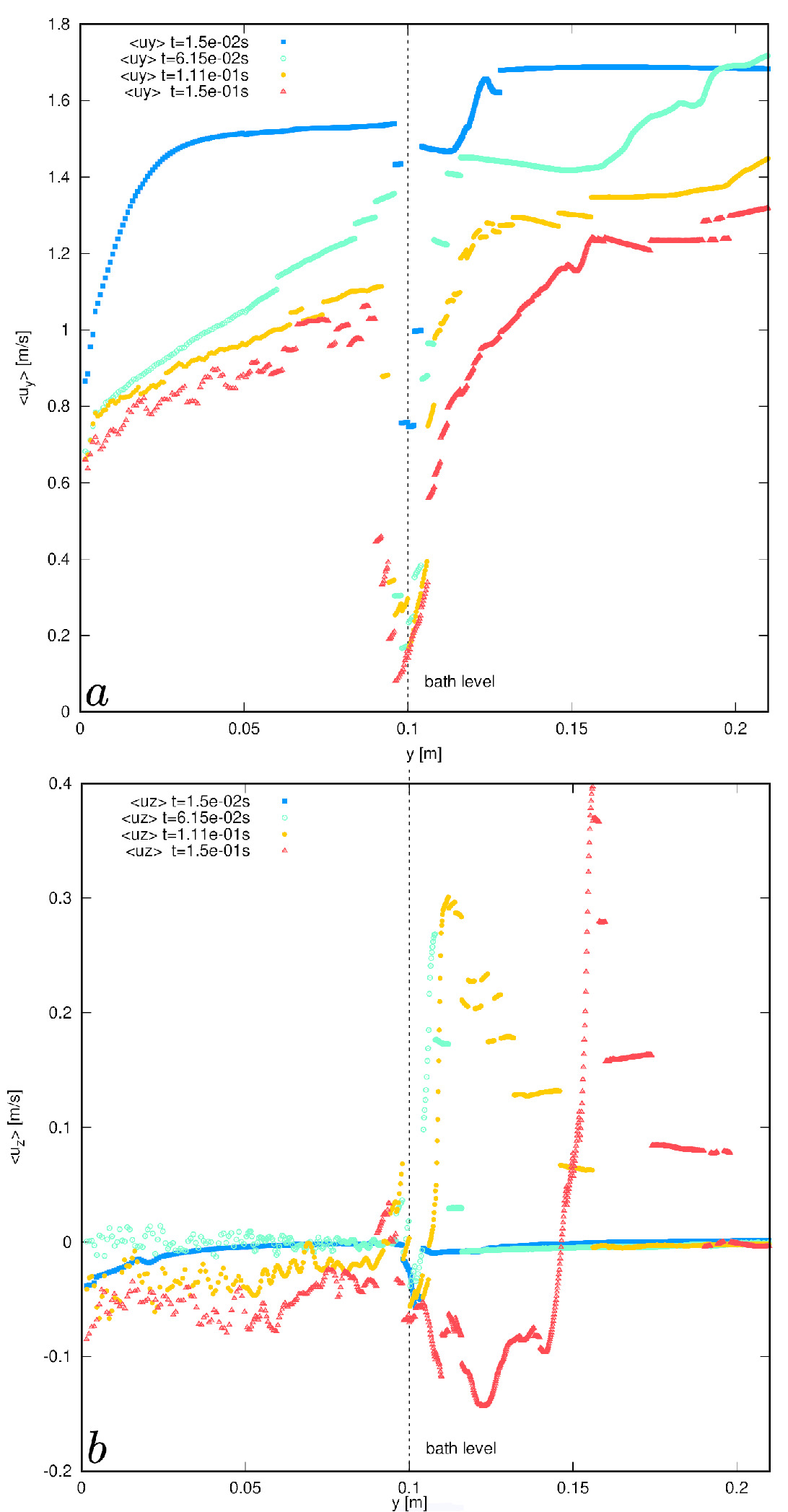}
  \caption{Coating in $G_{2,13}$ configuration: $z-$averaged velocity profiles  for $x\in \lb 0,0.001\rb$ at  varying $t$ values. (a) $u_y$ (vertical) velocity component; (b) $u_z$ (transverse) velocity component. }\label{g213prof}
\end{figure}

We continue our look at the physics of the three-dimensional film formation with Figure \ref{g213prof}, which contains velocity profiles for the vertical component ($u_y$) and the transverse component ($u_z$) along the wall height -- only the range of $y\in\lb 0,0.2\rb$ is included, as all $t$ values included are smaller than $t=0.15$s. At that time, the liquid reaches roughly to $y=0.3$m, consistent with Fig. \ref{2dints}. Note that profiles are $z-$averaged, and include data only for $x<0.001$m, thus the measurement window supplying data for Figure \ref{g213prof} includes only the closest proximity of the coated wall. In Fig. \ref{g213prof}a, we observe a transition from a rather smooth $u_y$ profile at $t=1.5\cdot 10^{-2}$s to a much more varied, at final pictured stages. Notably, we observe a stagnation region forming close to the bath level (itself drawn with a dashed line) which is consistent with interface geometry observed in Figure \ref{evogray2}. It is expected that $u_y<0$ velocities are present in this region further from the wall -- this however has not been captured with the profile measurement window. Average $\langle u_y\rangle_z$ values are consistent with Figure \ref{evogray2} as well (note that gas velocity is also taken into account in Fig. \ref{g213prof}). We now focus our attention on the curve for $\langle u_y \rangle_z$ at $t=0.15$s (red color in Fig. \ref{g213prof}a). This curve, although calculated using a three-dimensional simulation, is comparable with Fig. \ref{2dprof_thick} (curve for $t=0.1457$s). If, using the latter of the mentioned curves, one calculates a mean value (for $x\in\lb 0,1\rb$) of $u_y$, it is equal to $1.22$ m/s. This value should be at least comparable with Fig. \ref{g213prof}a taken for $t=0.15$s and $y=0.14$m; in fact, we find $\langle u_y\rangle_z (0.14)\approx 1.1$ which is within ten percent of the two-dimensional simulation. The slight discrepancy might be attributed to the $z-$averaging  in three dimensions; e.g. presence of the coated band edge, as well as wrinkles pictured in Fig. \ref{evogray2}.

Further evidence of the strictly three-dimensional character of the film is found in Figure \ref{g213prof}b, showcasing the profiles of the transverse velocity component $u_z$. While close to the beginning of the flow at $t=0.015$s (blue squares) this component is nearly zero, $u_z$ oscillates with increasing amplitude in the entrainment zone as time progresses, and remains negative everywhere below the bath level. That is to say the net flow of the liquid layers contacting the coated wall is \textit{from} the coated edge towards the symmetry plane (at $z=0$). As the film forms  and its top edge moves further from the bath, transverse  net flow is positive, i.e. \textit{towards} the coated edge, which is consistent with Fig. \ref{evogray2} and explains the rugged surface of the film in the edge neighborhood. Summarizing, it is obvious that at this $\Re_f$ values, three-dimensional effects are strongly present and decisive in determining the liquid flow character.

Finally, note also that Figure \ref{g213prof} features the most resolved of the 3D simulations presented in the paper, at $2^{13}$ which is locally equivalent to a grid of $8192^3$ points \footnote{Due to CPU time and memory restrictions we have not continued this simulation into the injection stages.}.

\subsection{Single-phase Impinging Jet Study}\label{YSsect}

A brief study has been performed on the velocity and pressure profiles obtained in boundary layers of a (two-dimensional), single-phase jet impinging on the flat plate. This is motivated by the need to ''calibrate'' the analytic predictions for the result of jet interaction with film. More precisely:  velocity, pressure and shear stress $\tau_{yy}$ profiles can be applied to extract coefficient when calculating the film thickness $h_c$ above the impact zone  \citep{tuu}. Note that in this 2D simulation, nozzle walls have been defined as slightly thicker ($2$mm instead of $1$mm used in the ''industrial'' configuration); this should have no effect on the pressure distribution in the air-wall impact zone.

\begin{figure}[ht!]
  \centering
  \includegraphics[width=0.42\textwidth]{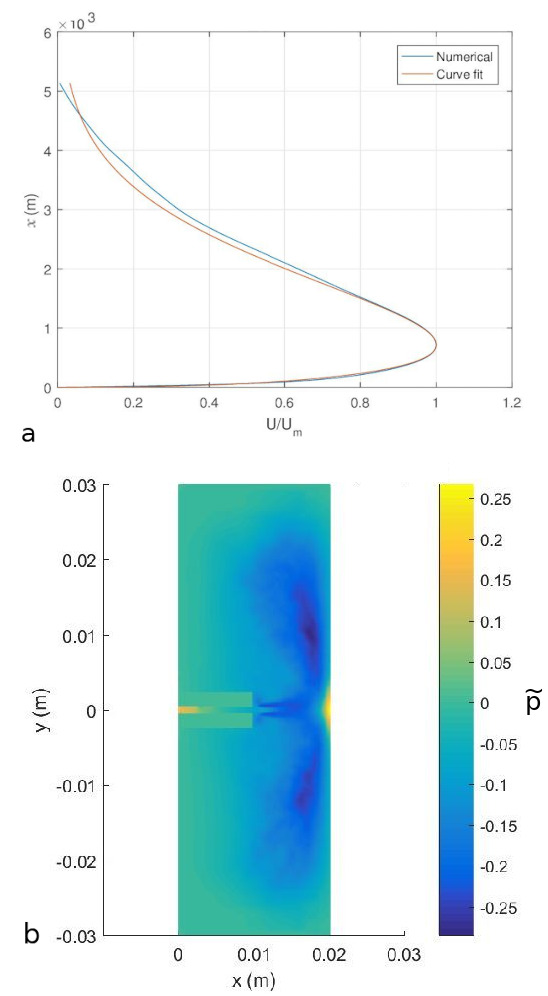}
  \caption{Study of an impinging jet (single-phase flow using two-dimensional $G_{2,11}$): (a) Velocity profile near the wall, simulation (ensemble averaged, blue) and \citep{ozdemir92} ((\ref{ozd92}) brown); (b) Ensemble-averaged mean pressure in the same simulation.}\label{ozdemir}
\end{figure}

To begin with, the velocity profile in the wall jet region  \citep{glh_nasa} has been extracted from the simulation using $u_{inj}=200$ms$^{-1}.$ We have applied a combination of time- and ensemble-averaging in order to ensure smoothness of profiles (in most situations 15 simulations were ensemble-averaged). Time-spans used for temporal averages were relatively large: of the order needed for largest vortices to leave the domain. In Figure \ref{ozdemir}a an example of wall velocity profile is presented.

Figure \ref{ozdemir}a presents the profile of the velocity component $u_y$ parallel to the wall (coated band in full simulation) normalized by the mean value $u_m,$ taken at $3\cdot d_{nf}/4$ distance from the stagnation point. The curve is accompanied by a fit to the analytic prediction presented by \citep{ozdemir92}, namely:

\begin{equation}
  \label{ozd92}
  \frac{u}{u_{max}}=\frac{\gamma}{\beta}\left(\frac{x/x_{0.5}}{\beta}\right)^{\gamma-1}\cdot\exp\left(-\left(\frac{x/x_{0.5}}{\beta}\right)^\gamma\right),
\end{equation}

where we assume $u_{max}=200$m/s as per problem specification, while coefficients $\gamma$ and $\beta$ are taken $1.32$ and $0.73$ respectively. Value of $x_{0.5}$ needs to be set such that it corresponds to the position in the outer layer where velocity is half of the maximum recorded between the two layers. Clearly, we observe in Figure \ref{ozdemir}a a boundary layer whose thickness is about $0.75$mm.

Figure \ref{ozdemir}b displays the spatial distribution of mean pressure $\tilde{p}$ normalized by the dynamic flow pressure,  or \[\tilde{p}=\frac{\langle p \rangle}{\frac{1}{2}\rho_g|\ub|^2}\] for the impingement simulation. One can observe correct symmetry in the distribution, as well as easily recoverable stagnation point directly opposing the nozzle outlet. Zooming into Figure \ref{ozdemir}b reveals that pressure changes sign very close to the wall, which is consistent with velocity curves predicted by (\ref{ozd92}) and the existence of boundary layer. According to \citep{tuu}, peak pressure evolves with the distance from the nozzle  as \begin{equation}\label{tuupres}\frac{p_s}{\frac{1}{2}\rho_g|\ub|^2} \approx \left(\frac{7d}{d_{nf}}\right)^{-1}.\end{equation} Values presented in Fig. \ref{ozdemir}b are about half of predicted by (\ref{tuupres}), meaning that the potential core is not resolved well enough for $G_{2,11}.$ This warrants an increase in refinement, and is one of the reasons for using at least $12$ levels of refinement in majority of the simulations.  Profiles presenting raw pressure values are shown below, e.g. in Figure \ref{synt_prof_p}a.

By fitting  the simplified $p_{air}$ and $\tau_{yy}$ curves, constructed using (\ref{nth8b}) to the results obtained in this section, we were able to establish the values of the $c_s$ and $c_p$ coefficients mentioned in \ref{nthsect}. For the industrial parameters, they amount to $c_p=0.09, c_s=0.00325.$ This results in the $\epsilon_3$ correction of (\ref{nth8}) equal to $0.461$. The above-nozzle film thickness $h_c$ is thus between $20$ and $40 \mu m$ for this configuration with thin layer assumptions. By using an iterative computation with the Colebrook-White  equation \citep{shashi} as is traditionally done in Moody  diagram applications \citep{moody}, we have estimated the wall friction coefficient for (\ref{nth11}) at $c_f\approx 1.48\cdot 10^{-3}$ resulting in a value for $\epsilon_5$ at $0.08,$  which is a correction associated with turbulent film regime; this increases the expected $h_c$ thickness (\ref{nth10}) above the injector by no more than $5-10 \mu m$  compared to the thin-layer approximated prediction. This small difference is in accordance with predictions of \citep{tharma} who have investigated a similar case of a rotating roll coating. They show that including or disregarding inertial effects changes the resulting thickness prediction only very slightly, especially for low capillary numbers ($\Ca<1$) which is the case in our work ($\Ca\approx 0.03$).

%
\subsection{Three-dimensional Wiping Simulations. Configuration $G_1.$}\label{fullmonty}

In this subsection, we present the first results of the three-dimensional simulations performed. This includes geometry specification given in Table \ref{Gtab} and Figure \ref{2d3dsetup}a.   This first presented simulation ran at $2^{12}$ refinement level; with $L_0=0.512$, translating to grid-cell size of

\begin{equation}
  \Delta x =\frac{0.512}{2^{12}} = 1.25 \cdot 10^{-4}m,\label{cellsize}
\end{equation}

or 8 cells in nozzle diameter $d$. This is not sufficient to resolve turbulent flow within the nozzle, nor the $h_c$ thickness, however Groenveld's thickness $h_{G}$ of $163\mu$m is in reach. Simulations of this (and higher) resolutions are possible using the  SRR technique described above; refinement level is decreased by as much as $5$-$6$ levels (or $64$ times) far from the region of interest, e.g. close to outer domain borders.  At the $2^{12}$-equivalent resolution, it is reasonable to expect liquid bulges and wrinkles whose thickness surpasses $h_{G};$ these should be captured in current simulation.

\begin{figure}[ht!]
  \centering
  \includegraphics[width=.4\textwidth]{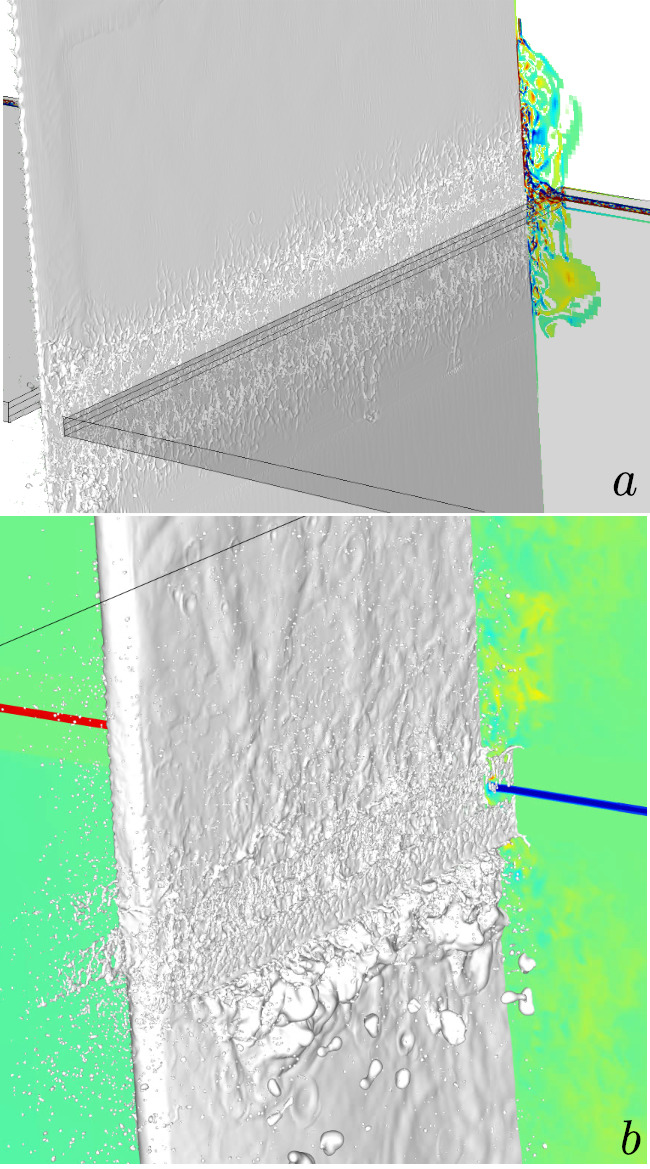}
  \caption{(a) $G_{1,12}$ with nozzle locations sketched as a shaded area. Color: $\omega\ne 0.$ (b) The $G_{1,12}(u_{inj}=75,u_{wall}=4)$ simulation in its final stages. }\label{prof3}
\end{figure}
To further examine the influence of the air-knives, we will proceed to three-dimensional, macroscopic visualization. An example is contained in Figure \ref{prof3}a which displays interface geometry at $t=0.163$s. Nozzle locations are drawn using shading and black outlines in this view -- this is done in post-processing and only for orientation purposes. Additionally, a cut-plane is positioned in the back-drop (parallel to  $z=0$ coordinate) colored by vorticity. At $t\approx 0.16,$ a relatively  wide  impact zone is already visible, with individual droplets ejected from the film, as well as rich wrinkling. The structure of the air trace is three-dimensional: even if its character is homogeneous above the nozzle, below it we see two zones with larger traces. Additionally, edge area is visibly atomized. Figure \ref{prof3}b emphasizes the consequences of certain geometrical differences between the industrial and relaxed parameter sets. Thicker coated plate is visible; the coat on the plate edge is seemingly not disturbed except in the impact area where it interacts directly with the turbulent structures resulting from collision of air that emanates from opposing nozzles. Moreover, some liquid deposits on the nozzle walls (the nozzles are not rendered in Fig. \ref{prof3}b) partly obscuring the view. Large amounts of the coating material crumble  down below the impact zone, resembling the ''peeling'' effect observed in \citep{myrillas}.

\begin{figure}[ht!]
  \centering
  \includegraphics[width=.48\textwidth]{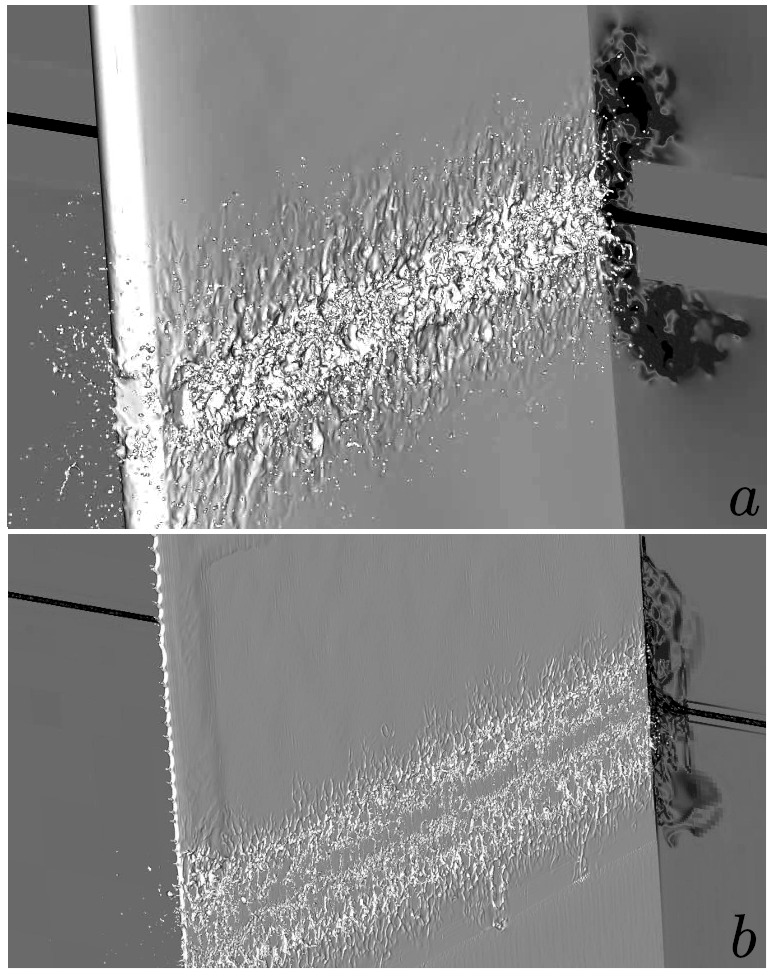}
  \caption{The $G_1$ simulation parameter study: (a) $G_{1,12}\left(u_{inj}=75,u_{wall}=4\right)$, (b) $G_{1,12}$ with industrial parameters.}\label{3dcompar}
\end{figure}
Another comparison of  relaxed and industrial parameters (as defined in Table \ref{realrelax}) is visible in Figure \ref{3dcompar}. Due to a difference in timescale, picture visible in Fig. \ref{3dcompar}a has been chosen based on the width of the impact zone. 

The main focus of the comparison is however the degree of film atomization which, in case of relaxed parameters, is visibly higher. This can be attributed to smaller density of the liquid phase, facilitating the momentum transfer from the gas. Also, film deposit is far thicker to the left ($z+$ direction), which, at the same resolution, means that zero-flux film is better represented in Fig. \ref{3dcompar}a than in b which is consistent with the \Re being much lower in the relaxed case: $5380$ (a) versus $14 300$ in (b) (``optimistic'' estimation based on nozzle diameter $d$) or $633$ (a) versus $2240$ (b) for the film (based on $h_{00}$ and $u_w$).

\begin{figure}[ht!]
  \centering
  \includegraphics[width=.5\textwidth]{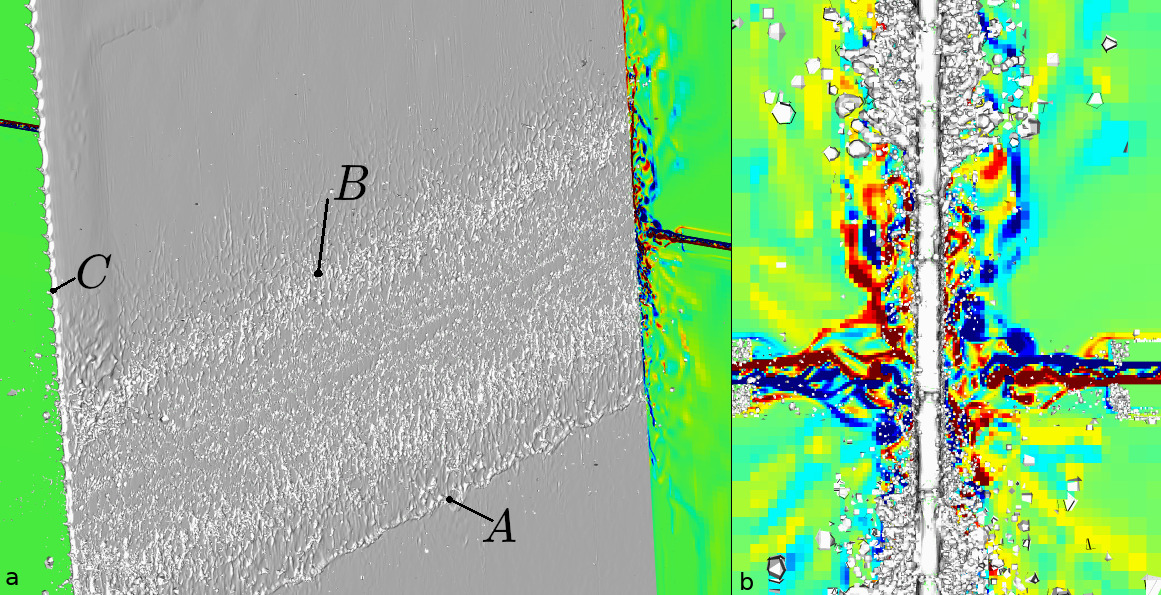}
  \caption{The $G_{1,12}$ simulation at $t=0.164$s. Views: (a) isometric and (b) side view (along $z$ axis). The back-drop cut-plane colored by vorticity.}\label{3dupdate}
\end{figure}

Continuing the flow analysis for the industrial case, we turn our attention to Figure \ref{3dupdate} which displays film geometry at approximately $0.17$s. By this time, the lower bulge (A) starts forming (below the nozzle level) leading to the onset of back-flow into the reservoir. It is visible at the side of the film as a distinct line in Fig. \ref{3dupdate}a. Basing the measurement on the profile of the fraction function $\langle C\rangle$, we estimate the width of the impact zone at $4$cm. By ''impact zone'' we understand an area (A-B in Fig. \ref{3dupdate}) with minimum $\langle C\rangle,$ not the entire area of gas/liquid interaction which, as visible in Figure \ref{3dupdate} is much larger: film shearing takes place in a nearly $0.2$m-wide area above and below the air-knives.

No significant edge effects are visible yet (C in Fig. \ref{3dupdate}, the wavy edge film structure is an early effect of film formation). However, inspecting Figure \ref{3dupdate}b we may conclude that the edge film is nearly entirely atomized. In Fig. \ref{3dupdate}b the same film is seen from a different viewpoint: along the $z$ axis, centered at the impact zone. A certain perspective shortcut effect takes place in Fig.\ref{3dupdate}, as  droplets close to the viewpoint, i.e. on the plate edge, seem bigger than those far from it. Besides, the view contains an apparent  accumulation of droplets from all plate depth (i.e. most droplets along $z$-depth of $0.15$m of the plate are visible) which seems to contradict Fig. \ref{3dupdate}a if the said perspective effect is not taken into account.

The total CPU cost of the 3D, $2^{12}$ simulation presented here is approximated at $122 000$ CPUh, only twice the amount of 2D simulations presented in previous subsections. However, maximum refinement level is four times lower here, and simulated physical time is only about one tenth that of the 2D simulation.

\subsection{The $G_2$ configuration results}\label{g2sect}

In this section, we present results pertaining to the $G_2$ configuration. As mentioned above, this configuration enables us to focus on the air-liquid impact study in more detail as, as long as the mean flow is considered, the $G_1$ configuration has an inherent symmetry. Placing the coated wall in the corner of a cubic domain, we use the SRR technique to coarsen the grid proportionally to the distance from the coated walls. This, as the simulations below confirm, has proven sufficient to dampen the turbulent flow far from the zone of interest and prevent backflows. %

  \begin{figure}[ht!]
    \centering
    \includegraphics[width=.48\textwidth]{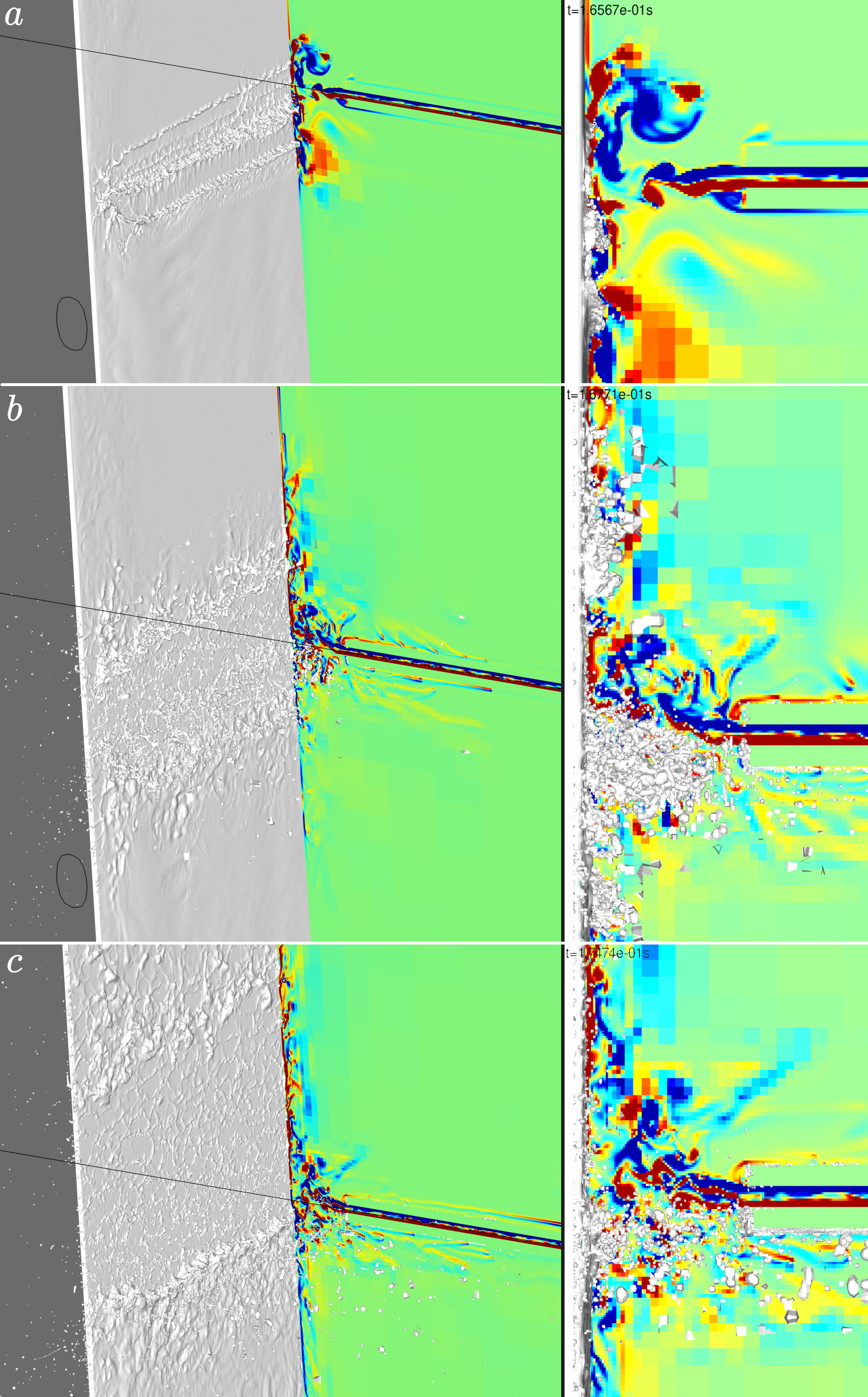}
    \caption{The $G_{2,12}\left(u_{inj}=200 \right)$ simulation at (a) $t=1.656\cdot 10^{-1}$s (b) $t=1.677\cdot 10^{-1}$s and (c) $t=1.747\cdot 10^{-1}$s. }\label{g2big}
  \end{figure}

  In Figure \ref{g2big} we present a visualisation of the macroscopic shape of the interface for the $G_2$ simulation performed using a $2^{12}$-equivalent grid.  This simulation corresponds to industrial parameters and is the $G_2$-analog of results mentioned above e.g. in Fig. \ref{3dupdate}. In three sub-figures, instantaneous shapes are visible for (a) $t=1.656\cdot 10^{-1}$s (b) $t=1.677\cdot 10^{-1}$s and (c) $t=1.747\cdot 10^{-1}$s. Each of the pictures presents two separate view: an isometric one on the left-hand-side, and a side-view (looking along $z$ axis) on the right-hand-side. The cut-plane positioned at the $z-$ domain wall is colored with $\omega.$ Varying cell size in the vorticity cut-plane is, of course, a consequence of employing the aforementioned SRR technique to limit adaptivity in regions above and below the nozzle. Full resolution is maintained inside the planar nozzle and  within one mm of the coated wall. As visible in the r.h.s. images of Fig.\ref{g2big}b and Fig.\ref{g2big}c, the grid coarsening affects interfacial formations as well: the ejected droplets and ligaments are represented with a coarser grid the further from the coated wall.

  Directly after the air contacts liquid in Fig.\ref{g2big}a, we note a distinct imprint of the nozzle shape on the liquid. Three longitudinal bulges are formed: one below the nozzle, one directly opposing the air outlet and lastly, a small bulge is formed above the nozzle. A mere two milliseconds later, as shown in Fig.\ref{g2big}b, the central bulge - whose liquid is ''trapped'' by the airflow, has been completely atomized, turning it into a cloud of droplets and ligaments. (This is shown particularly well in the side-view.) This last result is consistent with that of \citep{yu14}, who have (in 2D) investigated a flow characterized by a higher $\We$ of $13.5$ with lower density ratios. Their results show a $\langle C\rangle$ distribution consistent with a cloud of droplets -- with temporal averaging, it is displayed as a bulge.

  The atomization process results in most of the liquid droplets being rejected out of the field of view. Some examples of fast-moving ``glider'' droplets are visible as traces just below the nozzle in Fig. \ref{g2big}b and c. In the meantime, the lower and upper bulges move away from the nozzle. In Fig. \ref{g2big}c, we note that the upper bulge has, by $t=1.747\cdot 10^{-1}$s advanced approx. $10$mm upwards, and has been considerably smoothed. Compared to the lower bulge, there is almost no atomized material near the upper one. Meanwhile, as suggested by the right-hand-side view in Fig.\ref{g2big}c, the material below the nozzle is partly stripped from the wall and immediately atomized. Fig. \ref{g2big}b suggests that most of the droplets in the impact zone originate from the atomized middle bulge material. Subsequently, the number of droplets below the nozzle in Fig. \ref{g2big}c is far smaller than visible in Fig.\ref{g2big}b suggesting that atomization visible in Fig. \ref{g2big}b is a transient phenomenon. 

  Above the level of the nozzle and between the bulges, a thin film is formed, covered by a three-dimensional wave structure as visible in Fig. \ref{g2big}c. At this resolution, we have $\Delta x = 125 \mu m$ which, compared with the prediction for $h_c\approx 40 \mu m$ (using thin film approximation (\ref{nth8})) or up to $50 \mu m$ (using (\ref{nth10}) means that thickness is about half of $\Delta x$. Fortunately, within the Volume of Fluid framework, maintaining cells that hold a film thinner than $\Delta x$ is possible \citep{tsz}, as well as advecting such structures.  Thus, we conclude that in certain regions of the wall the thinner film is correctly represented.

  Atomization of the film occurring at the first instance the air-liquid contact might be investigated looking at the non-dimensional numbers characterizing this interaction. While the film Reynolds number $\Re_f$ characterizes mostly film formation, we formulate the Weber number $\We$ involving gas velocity, as follows:

  \begin{equation}\label{weber}
    \We = \frac{\rho_g \ub_g^2 h}{\sigma},
  \end{equation}

  with $h$ standing for film thickness. Definition (\ref{weber}) is first applied to industrial parameters characterized by $u_{inj}=200.$

  \begin{table}[h!]
  \centering
  \begin{small}
    \begin{tabular}{ l c c c c  }
      \hline
                      &    $h_{00}$      & $h_G$          &  $h_c$                     & $h_{c,turb}$       \\
      \hline
      Value           &    \num{5.46e-4} & \num{1.63e-4}  &  $\approx$\num{3e-05}      & $\approx$\num{3.7e-5}       \\
      $\Re_f(\cdot)$  &   \num{2.24e03}  & \num{6.72e02}  &  \num{1.23e02}              & \num{1.52e02}               \\
      $\We(\cdot)$    &   \num{3.8e01}   & \num{1.14e01}   &  \num{2.09e00}             & \num{2.58e00}              \\
    \end{tabular}
    \caption{Values of film thickness and the resulting dimensionless numbers for the industrial air-knife configuration. Values for $\Re$ and $\We$ are calculated using the average values of $h_c$ and $h_{c,turb}$.}
    \label{thicktab}
  \end{small}
\end{table}

  Feeding the zero-flux thickness  (\ref{h00eq}) into (\ref{weber}) one obtains $\We(h_{00})=38.1.$ If, instead, we settle upon using Groenveld's thickness $h_G,$ (\ref{weber}) yields $\We=11.4.$ Both these values seem consistent with a regime in which atomization might be expected \footnote{For the ``relaxed'' configuration presented previously, using $h_{00}$ is more justified as the film Reynolds number is three times lower. Doing so, we obtain $\We(h_{00})=7.57.$ }. Values of film thickness calculated using various definitions are given in Table \ref{thicktab}, which contains also resulting values of the film Reynolds number as well as Weber number.  As  the atomization effect has not been reported previously  \citep{myrillas, ellenvu} we have decided to study it further. This is motivated by the fact that similar liquid breakup could be induced numerically, e.g. by inconsistent momentum transfers  \citep{geo2015}, curvature calculation errors, or not accounting for interactions between liquid-gas interface and vortical structures in the latter phase  \citep{aniszewskiJCP2016}. Thus, we have included simulation configured as $G_{2,12}(u_{inj}=42)$ by decreasing air injection velocity. This configuration is characterized by $\We(h_{00})=1.68$ and $\We(h_G)=0.5$ which, again, places the system just below the ''edge of criticality'' as in a context of $\Re_f$ in our film formation study.   We follow up with an examination of the flow at a decreased injection velocity. 

  \begin{figure}[ht!]
    \centering
    \includegraphics[width=.5\textwidth]{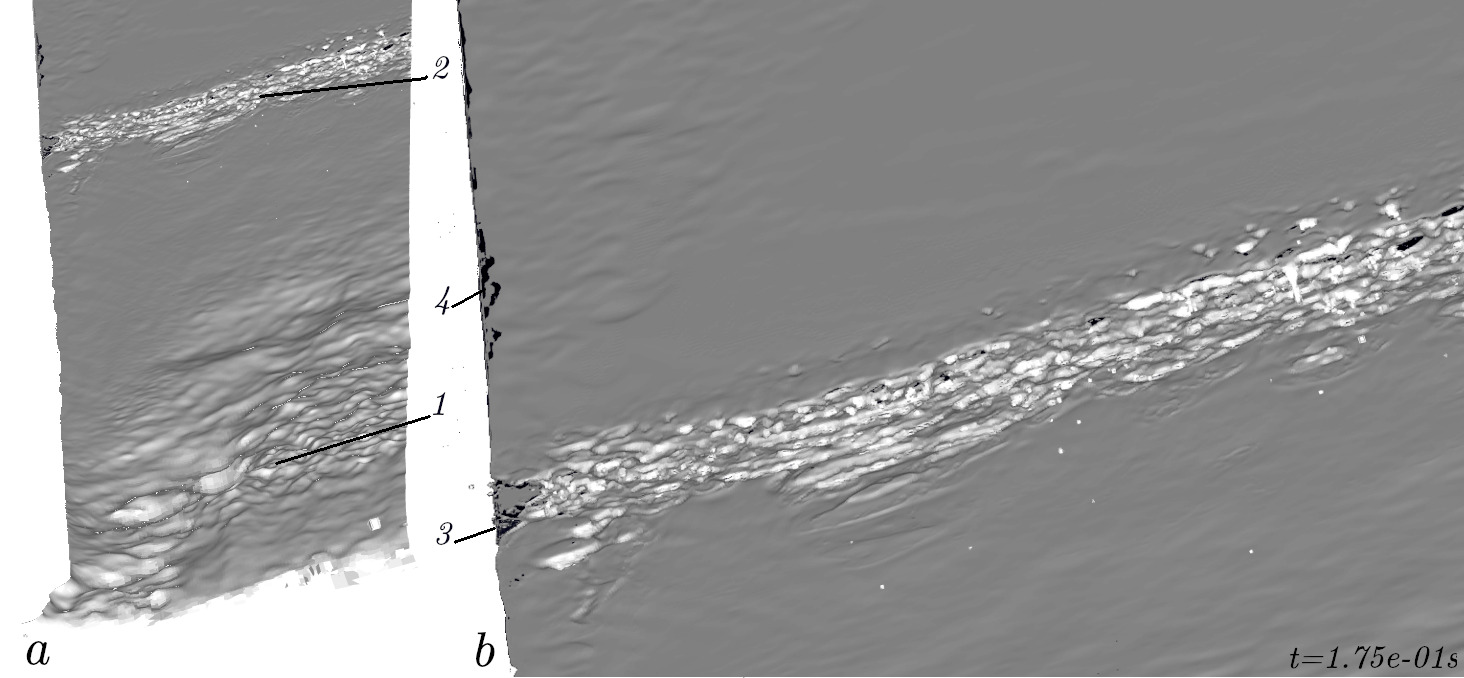}
    \caption{The $G_{2,12}\left(u_{inj}=42 \right)$ simulation at $t=1.756\cdot 10^{-1}$s. (a) Isometric view;  (b) zoom into the impact region at the same time instant. Navy blue color indicates the  wall (where coating is absent). $1:$ bulges created in the high $\Re_f$ withdrawal, $2:$ air impact area, $3:$ coating defect in the impact zone $2$, $4:$ coating defect above the impact zone. }\label{rtis42iso}
  \end{figure}

  Figure \ref{rtis42iso} presents the $G_{2,12}(u_{inj}=42)$ simulation roughly $2$ milliseconds  after air-liquid impact. In the Figure, white shaded isosurface represents the liquid interface, while several blue areas depict the (un-coated) moving wall\footnote{In Fig.\ref{rtis42iso}a, the bath level is over-exposed  (i.e. rendered as white) due to specific light positions in the visualization.}. The far view presented in Fig. \ref{rtis42iso}a confirms again the turbulent film character below the impact zone (denoted ''1'' in the Figure). Interface geometry in the entrainment region 1 is comparable to Figure \ref{evogray2}, and should not be associated with the air-liquid interaction. The impact zone is visible above as an area with horizontal wrinkles (Fig. \ref{rtis42iso}a:2 and Fig. \ref{rtis42iso}b). Looking closer at the impact zone we note a small number of gaps (denoted 3) in the film, mainly close to the edges  of the coated band. Defects may results from  the expected film thickness being  not fully resolved. There are however visible edge coating defects (denoted 4 in Fig. \ref{rtis42iso}b) not likely associated with airflow. This is consistent with Fig. \ref{evogray2} and seems to suggest that not only increased resolutions are required in the neighbourhood of the coated edge, but possibly specific formulation of boundary conditions at the sharp solid edge (singularity).

Regarding the  atomization phenomenon at instance of air-liquid contact, comparing Fig. \ref{rtis42iso} with right-hand-side images in Fig. \ref{g2big} we note the nearly complete lack of atomized structures for $u_{inj}=42$m/s and $\We \approx 1.$

  \begin{figure}[ht!]
    \centering
    \includegraphics[width=.48\textwidth]{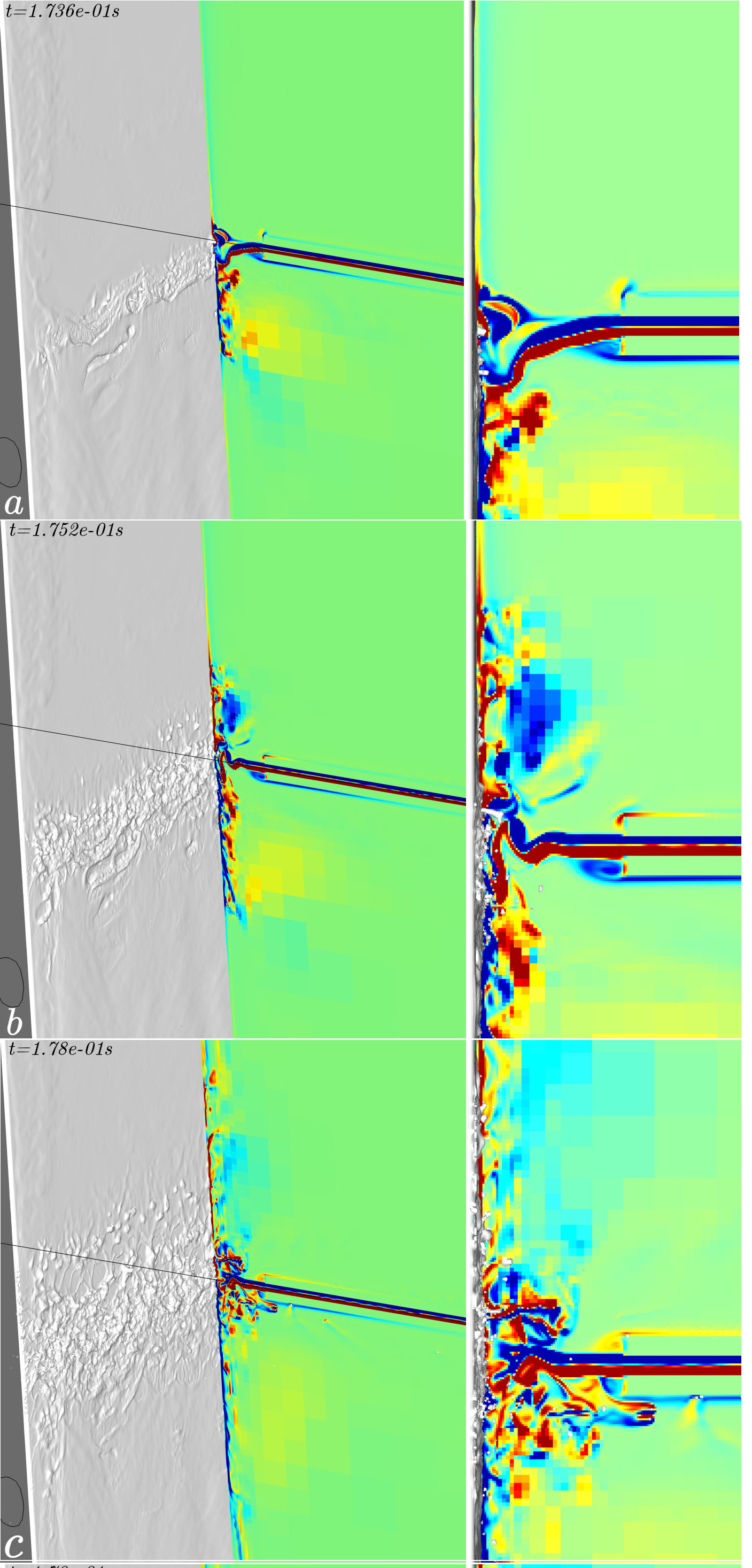}
    \caption{The $G_{2,12}\left(u_{inj}=42 \right)$ simulation at: (a) $t=1.736\cdot 10^{-1}$s, (b) $t=1.752\cdot 10^{-1}$s,(c) $t=1.78\cdot 10^{-1}$s. }\label{rtis42big}
  \end{figure}

 Another look at the flow characterized by lower Weber number is provided in  Figure \ref{rtis42big}. Three sub-figures present the same flow as pictured in Figure \ref{rtis42iso} at instances of time with $t\in\lb 0.1736,0.178\rb$. Figure \ref{rtis42big}a presents interface geometry in the impact zone almost directly following the first air-liquid contact\footnote{Note that Figure \ref{rtis42iso} corresponds to the instance of time situated between Fig.\ref{rtis42big}b and c.}. Differences caused by the decrease in the Weber number are instantly recognizable: no distinct horizontal liquid bulges are formed along the $z$ direction; instead, smaller-wavelength disturbances are showing within a gradually broadening region, as seen  in Fig. \ref{rtis42big}c. The side-views included in the Figure show a significant decrease in the number of droplets, which we quantify below in Figure \ref{dropscom}. Overall image of the flow is different than that for $\We>10$. Juxtaposing Fig. \ref{g2big}b with Fig. \ref{rtis42big}c  we note the complete absence of the droplet cloud below the impact area. We suspect that in the low-$\We$ regime the film is merely disturbed by the airflow, while areas above the nozzle are continuously fed liquid; hence no permanent film thinning should be expected in such flows. 

\begin{figure*}[ht!]
    \centering
    \includegraphics[width=\textwidth]{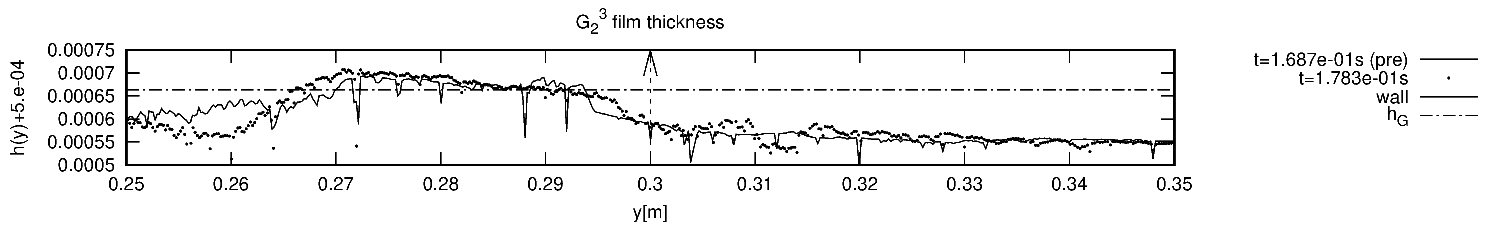}
    \caption{Film thickness profiles for the $G_{2,12}(u_{inj}=42)$ simulation. }\label{rtis42prof}
\end{figure*}

Indeed, looking at the film thickness profiles presented in Figure \ref{rtis42prof}, we note that the $z-$averaged film thickness in the vicinity of the impact zone (the nozzle level is marked with an arrow around $y=0.3$m) has been altered but not significantly diminished, except the area some $15$mm above the nozzle. In the Figure, $h(y)$ profiles are shown for two instances: $t=1.687\cdot 10^{-1}$ (continuous line) and $t=1.783\cdot 10^{-1}$ (black dots). Groenveld's thickness $h_G$ is drawn in dashed line for comparison purposes.  Profiles visible in Fig. \ref{rtis42prof} show clearly a bulge for $y\in\lb 0.265,0.295\rb$. This formation can not be simply associated with the jet influence, as it is present as well in the profile prior to impact; it is more likely that it results from uneven coating. At this height, the film is characterized by dimples \citep{snoe} -- visible in Fig. \ref{rtis42big}a, also visible in 2D in Fig. \ref{2dints} above $y=0.3$m --  and the coat is of three-dimensional character, being on average thicker closer to the $z+$ edge. This is thus displayed in the profiles below the impact zone. As for the consequences of the impact itself, $h(y)$ oscillates in the vicinity of $y=0.3$ which is fully consistent with instantaneous images in Fig. \ref{rtis42big}a-c and indicates alternating areas of thinning and thickening of the film. (Note that the zero-level shift in Figure \ref{rtis42prof} is a correction for the wall thickness of $5\cdot 10^{-4}$m.)  By representing $p_{air}(y)$ and $\tau_{yy}$ using gas velocity as in (\ref{nth8b}), we note their magnitudes for $u_{inj}=42$ are one order below that for $u_{inj}=200,$ which in the context of (\ref{nth8b}) and (\ref{nth8}) amounts to higher $h_c$. While at first sight it would seem consistent with results presented in Fig. \ref{rtis42big} and Fig. \ref{rtis42prof}, a far longer simulation would be required to establish the actual post-impact film thickness $h_c$ (by widening the area in which thinner film would be established) which was not the objective of this parameter study. As mentioned, thickness is diminished for $y\in \lb 0.31,0.32\rb$, with the thinner area coasted by two slight bulges. These formations are visible in Fig.\ref{rtis42big} as horizontal sets of dimples above the impact zone. In our opinion, both the film thinning for $y\approx 0.314$ as well as the bulges are artifacts of the collision of a large horizontal vortical structure with the film. Similar effect should be observed in a longer timescale. Namely, individual, spatially distinct ''craters'' are probably created on the film surface by individual vortical structures - separated by distances resulting from the jet flapping frequency. 
 
\begin{figure}[ht!]
    \centering
    \includegraphics[width=0.5\textwidth]{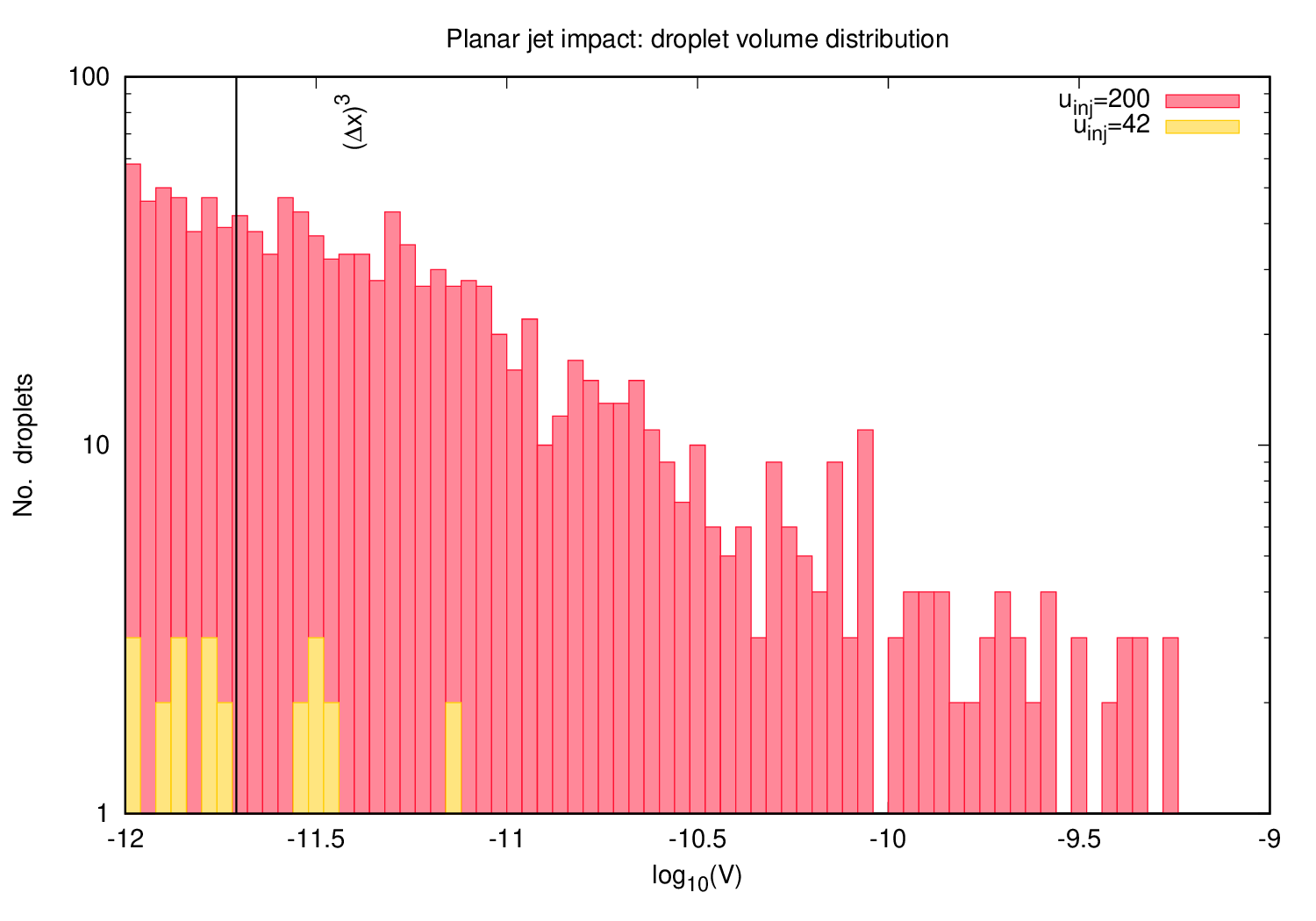}
    \caption{Droplet size distribution in the $G_2$ configuration for varying air injection density.}\label{dropscom}
  \end{figure}

Figure \ref{dropscom} presents droplet volume distribution for $G_{2,12}(u_{inj}=200)$ (pink bars) and $G_{2,12}(u_{inj}=42)$ (yellow bars). In the Figure, minimum cell volume ($\left(\Delta x\right)^3$ at the finest grid level) is denoted with a black vertical line. Clearly, both simulations involve a significant number of 'sub-grid' VOF ''debris'' -- grid-cells containing non-zero fraction function values that cannot be geometrically reconstructed. This is due to the fact that, firstly, turbulent airflow contributes to droplet breakup which continues until grid resolution becomes insufficient. Secondly, the SRR technique makes this mechanism act much more often which can be seen e.g. in the r.h.s. image of Figure \ref{g2big}c. Focusing our attention on the resolved droplets, we note in Fig. \ref{dropscom} that at low injection velocity there is about $15$ resolved droplets in total (yellow bars) which is qualitatively different than at higher air velocity (red bars).

\begin{figure}[ht!]
    \centering
    \includegraphics[width=0.5\textwidth]{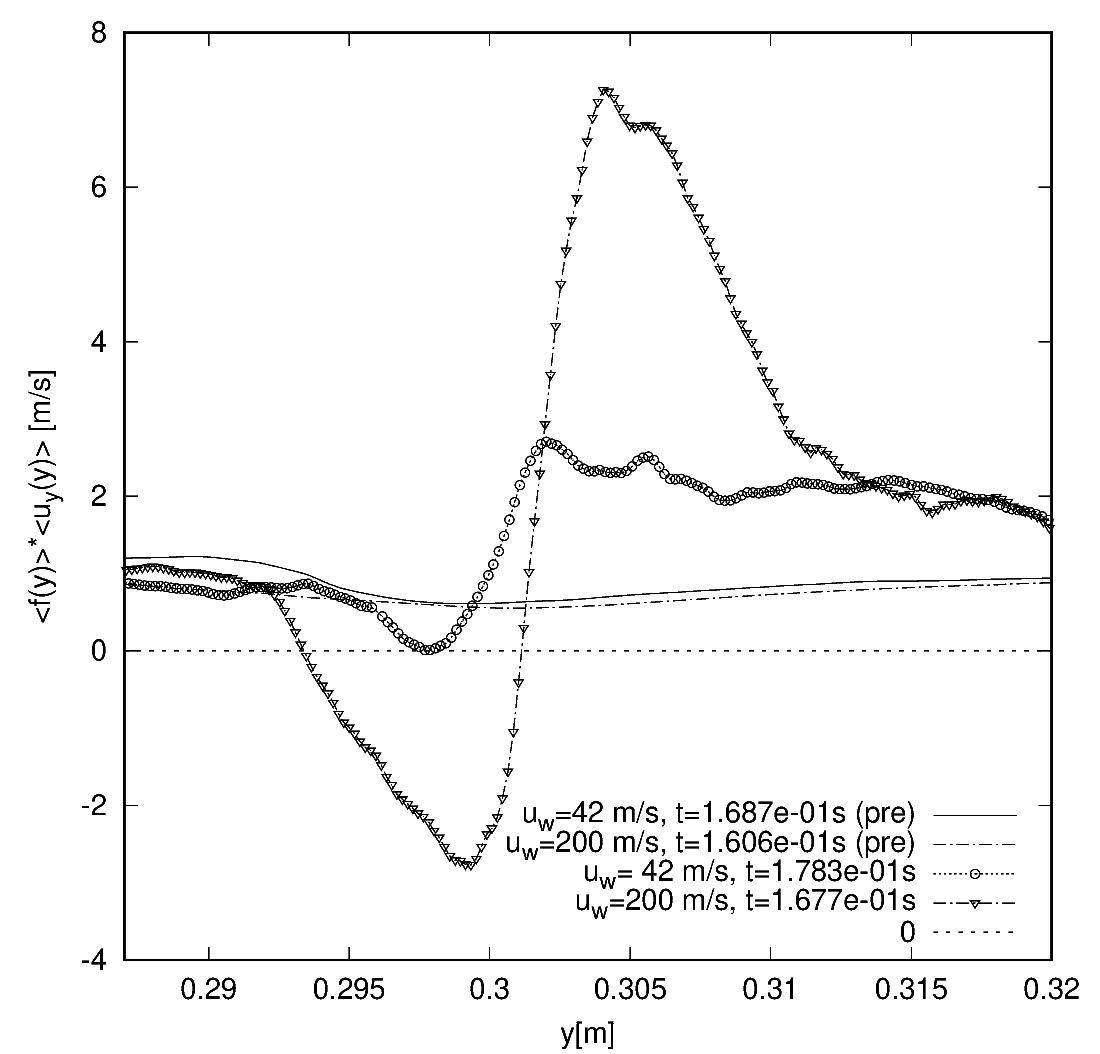}
    \caption{Average liquid velocities ($u_y$ component) in the impact zone.}\label{velcomp}
\end{figure}

An attempt to characterize the influence of impinging gas flow onto the internal velocities of the liquid film is presented in Figure \ref{velcomp}. In the Figure, we are looking at the approximated vertical component of liquid velocity obtained using the VOF fraction function $C$, which is equal to $1$ inside the liquid. In other words the product $C(\xb)u_y$ disappears in the gas phase, and Fig. \ref{velcomp} shows its profile in the direct neighbourhood of the impact zone ($y \in \lb 0.28, 0.32\rb)$ i.e. two centimeters below and above the impact zone). Two simulations are included with $u_{inj}=42$ and $200$ m/s. Since two flows have slightly different characteristic time scales due to higher $\We$ in the later, we have compared instantaneous profiles at the instance corresponding to the impact zone width approximately equal $0.01$m (as e.g. in Fig. \ref{rtis42big}c). The curves have been averaged over a sampling window $1$mm thick. For both flows, pre-impact velocity distribution in the analysing window is similar with $\lb u_y \approx 1$, which is caused by a film thinning in the analysing window slightly below the impact zone (i.e. the film is not perfectly flat even before the impact). After the impact, in case of the high-$\We$ simulation (inverted triangles in Fig. \ref{velcomp} one immediately observes the downward flow caused by the gas in the impact zone. Strong upward movement is visible above it. In the case of lower $u_{inj}$ the average $u_y$ values remain positive, suggesting the air knife wiping is far weaker for chosen injection parameters. Note that the overall character of the profile curves drawn using the averaging chosen in Fig. \ref{velcomp} remains comparable to ``raw'' velocity profiles as presented in Fig. \ref{synt_prof_ab}.

This concludes our investigation of the influence of the Weber number on atomization process - we conclude that the atomization effect visible at higher $\We$ is a correct result. The simulated air-liquid system responds as expected to the decrease in dimensional number, while other simulation parameters (e.g. grid resolution) are kept constant.

  We finish our analysis of the $G_2$ simulations with a brief remark on the   computational efficiency. Thus, the  approximated computational cost for the $G_{2,12}$ \textit{Basilisk} simulations presented e.g. in Figure \ref{dropscom}  was  $4.67\cdot 10^5$ CPU-hours (for simulation with $u_{inj}=42$m/s) and $4.8\cdot 10^5$ CPU-hours (for $u_{inj}=200$m/s simulation).
  
\subsection{The $G_3$ configuration results}\label{g3sect}

The $G_3$ configuration includes a geometry corresponding to a subset of the $G_2$ configuration (like the latter, which itself is a subset of $G_1$. This time only the area directly adjacent to the central section of the impact zone is included, as the domain size $L_0$ is ten percent that of $G_2.$ In this way we are able to resolve e.g. the flow within the nozzle even at refinement levels lower than that used while solving $G_2$ discussed previously. For example at $11$ levels of refinement, we have \[d=0.001=40\Delta x\] which is sufficient even for a rudimentary representation of the boundary layer within the nozzle. The boundary conditions however do not involve the liquid metal bath. Instead, the film is predefined at chosen thickness; for the results presented in this subsection $h_{00}$ thickness is chosen. Resulting film thickness will thus depend only on air knife pressure and shear \citep{hocking}, as there is no flux through the film. Simulations presented in the subsection have been carried out for $\We\left( h_{00}\right)=1.68$. Thus, some film wiping is expected to take place after the air-liquid interaction, however the coating will likely not be fully removed. To put the simulations into perspective: a thicker film ($h_{00}>h_G$) is included in the $G_3$ configuration, with $u_{inj}=42$ that is similar to ones used in Fig. \ref{rtis42iso} and \ref{rtis42big}.

  \begin{figure}[ht!]
    \centering
    \includegraphics[width=0.3\textwidth]{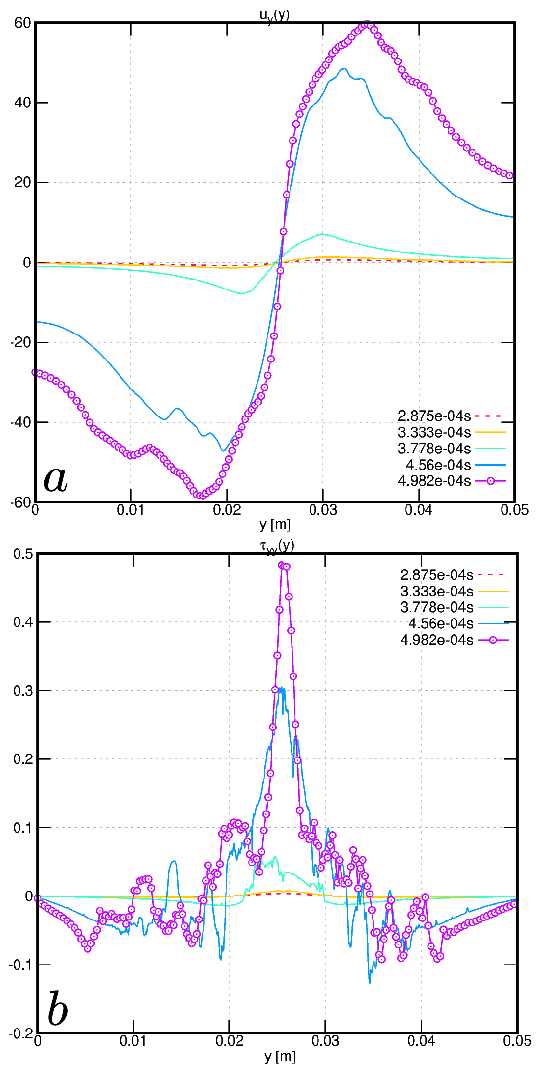}
    \caption{$G_{3,11}$: (a) $u_y$ velocity component (b) $y-$component of shear $\tau_{yy}$ at varying moments in time.}\label{synt_prof_ab}
  \end{figure}

  Figure \ref{synt_prof_ab} presents example profiles for the  $G_3$ type simulation with $L_0=0.0512.$ In this simulation, gravitational coating of the wall was replaced with a pre-defined film of thickness $h_{00}.$ Plots are instantaneous, which explains profile roughness; ensemble averaging, due to high associated cost, is not a viable option for this and other configurations.
  Symmetry of the vertical velocity component in Fig. \ref{synt_prof_ab}a indicates that air is evenly distributed upwards and downwards of the injection zone -- meaning that no jet flapping has occurred yet at $t=4.982\cdot 10^{-4}s$, which is $t_{inj}+0.1$ms. This, as shown below, is consistent with film profiles taken at that same instant in time. The lack of flapping phenomenon seems confirmed by the symmetric $\tau_{yy}$ distribution visible in Fig. \ref{synt_prof_ab}b. 

    \begin{figure}[ht!]
    \centering
    \includegraphics[width=.38\textwidth]{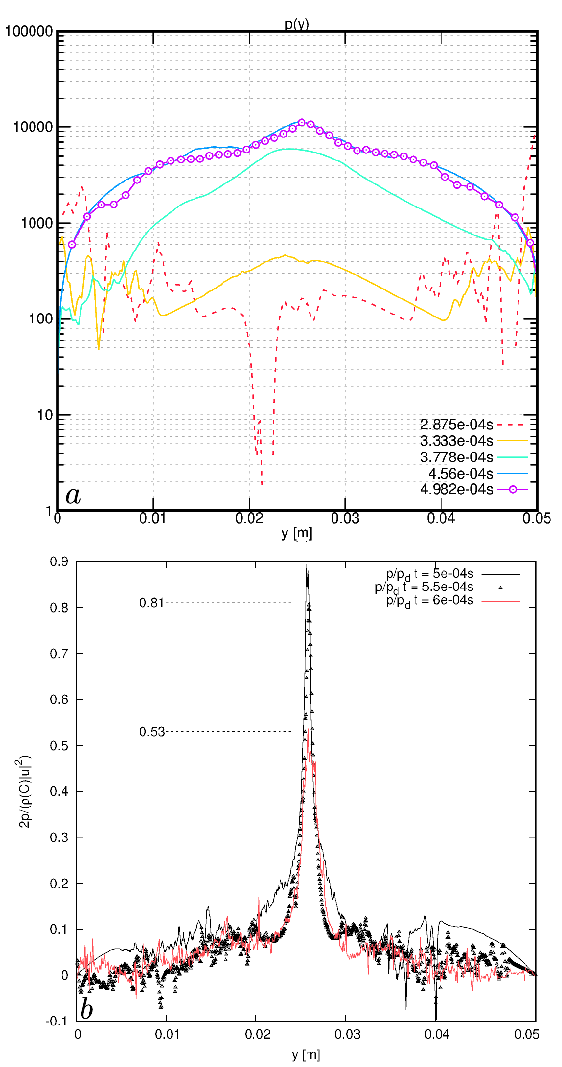}
    \caption{$G_{3,11}$: Pressure profiles.}\label{synt_prof_p}
  \end{figure}

    Profiles presented in Fig. \ref{synt_prof_ab} have however been averaged spatially, namely they are  sampled using a window which has a one millimeter span in $x$ and spans the whole domain in the $z$-direction. The same applies to pressure profiles presented in Figure \ref{synt_prof_p}. Logarithmic  $\langle p\rangle_z$ profiles visible in Fig. \ref{synt_prof_p}a exhibit a visible progression from the moment of air-liquid contact roughly at $t=3.33\cdot 10^{-4}s.$ Total pressure growth is suppressed after $5\cdot 10^{-4}s$, thus we turn our attention to a $q$-normalized quantity $\langle\tilde{p}\rangle_z$ in Fig. \ref{synt_prof_p}b, similarly to what we have done in the ``air-only'' study presented in Fig. \ref{ozdemir}. A decrease is visible converging close to the value of $0.6$ predicted by \citep{tuu}. Note that density used to prepare Fig. \ref{synt_prof_p}b is fraction function-weighted (effectively density-weighted) , namely \begin{equation}\label{vofdens}\rho=C(\xb)\rho_l+(1-C(\xb))\rho_g,\end{equation} which is a necessary step to include the phase mixture presence in the sampling window. Improvement in the $\langle\tilde{p}\rangle_z$ profile are likely the effect of a much higher relative resolution within the nozzle. Basing on  the turbulence characteristics and visualizations obtained in simulations for $G_3$ configuration (not shown), we are concluding that the results are not yet fully converged and substantially longer simulations are needed in order to make sure turbulent channel flow \citep{adams99} is approximated correctly. 

    \begin{figure*}[ht!]
      \centering
      \includegraphics[width=\textwidth]{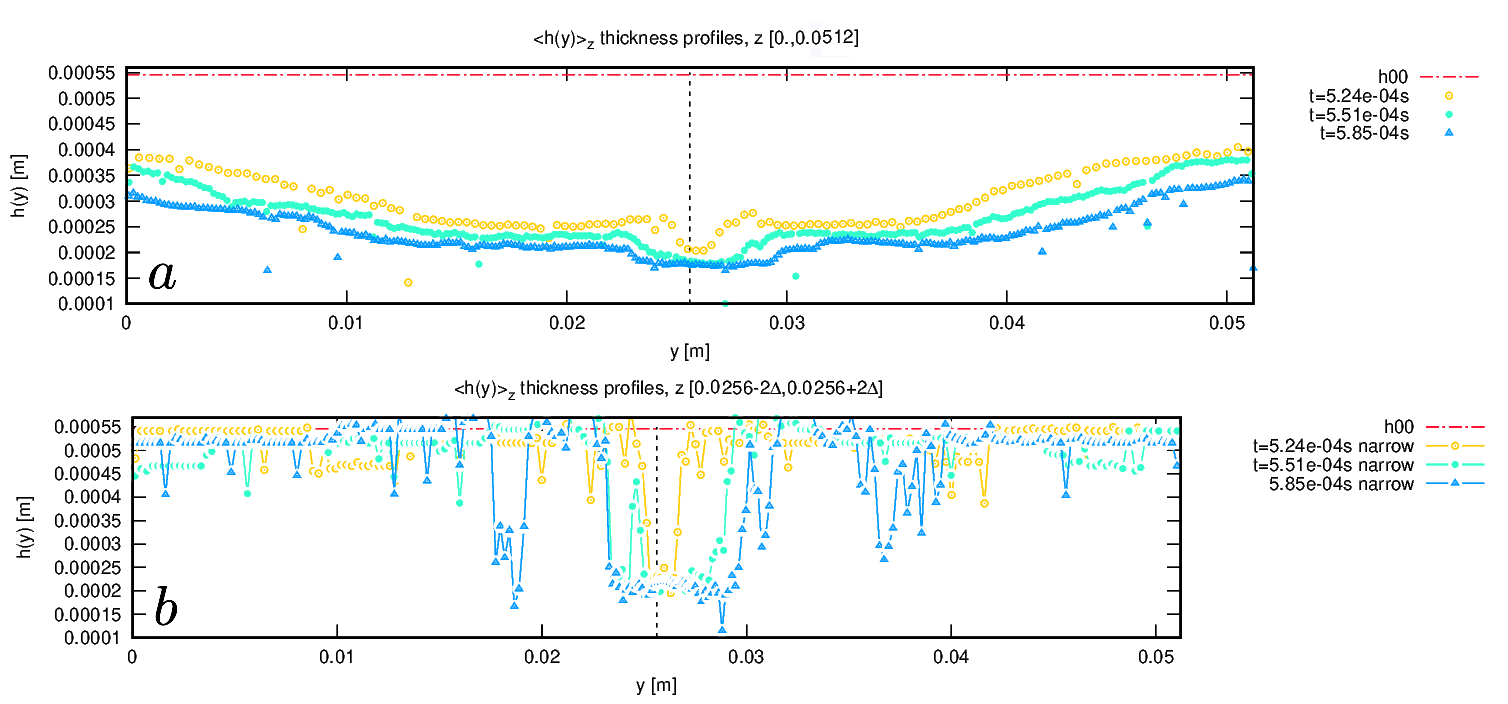}
      \caption{$G_{3,11}$: Film thickness profiles for an example $G_{3,11}(u_{inj}=42)$ simulation: (a) sampled using entire $z-$span; (b) using only a narrow strip around $z=0.0256.$}\label{synt_prof_th}
    \end{figure*}

    These suspicions are substantiated in Figure \ref{synt_prof_th}, showcasing the film thickness profiles for the $G_3$ type simulation. Two groups of profiles are shown: instantaneous profiles sampled along $y$ axis using the entire depth ($z$-span, or $z\in \lb 0,0.05\rb$) in Fig. \ref{synt_prof_th}a, and profiles sampled using only a narrow strip in the middle of the film or $|z-0.0256|\le 2\Delta$ in Fig. \ref{synt_prof_th}b.  In both subplots of Fig. \ref{synt_prof_th} the zero-flux thickness $h_{00}$ has been marked using a dot-dashed line.

    In is important to note that in the $G_3$ configurations film thickness is pre-defined, as $f_g=0$ and the coating process is not included. Thus, since $h_{00}\approx 3.36h_G,$ film thickness before impact is artificially kept higher than in $G_1$ and $G_2$ configurations, as the latter use gravitational coating. This decision is motivated by the decreased computational cost associated with representing a film of predefined thickness; however simulations using $h(y,t=0)=h_G$ are planned for future research.

    Returning to Figure \ref{synt_prof_th}, one observes that narrow profiles in Fig. \ref{synt_prof_th}b exhibit thickness close to the initial thickness $h_{00}$ along entire $y$-span except the centre of the impact zone. At the same time, profiles taken along entire $z-$span (Fig. \ref{synt_prof_th}a) describe a visibly thinner film. This, along with the analysis of macroscopic film geometry, confirms that the film is not thinned evenly along the entire $z-$span of the domain. In other words, a $z-$ stratification appears in the nozzle flow suggesting higher mean velocities close to the domain edges. Similar behavior might be observed in a channel flow which is not fully developed, and  small scale turbulent structures are mostly present close to domain edges. However, it seems that in this particular case film thinning occurred mostly close to the edges and did not take place in a manner similar to $G_1$ and $G_2$ configurations. This difference, in the context of this paper, should be explained by much lower spatial resolution of the $G_1$ and $G_2$ simulations compared to $G_3$ ($8$ points in $d$ for $G_2$ versus $40$ for $G_3$). For the continued research, inlet conditions for the air in $G_3$ configurations will be generated in such way that fully turbulent channel flow is obtained throughout the nozzle.

    Inspecting Figure \ref{synt_prof_th}b we find the film thickness in the impact zone at approximately $y\in\lb 0.025,0.03\rb \Rightarrow h(y)\approx 170\mu m.$ As mentioned above regarding the $G_2$ configuration, $h_c$ values up to order of magnitude higher than those expected at $u_{inj}=200$ are expected at this injection velocity. However, a more involved analysis of the $G_3$ configuration, which is ''synthetic'' in that it doesn't involve gravitational film formation (or $\mathbf{f}_g=0$), makes it clear than the analysis presented in Sect. \ref{nthsect} doesn't apply to this particular case. For example, $q=0$ as $u_{wall}=0$ and only $h_c=0$ could be accepted as a ''steady state'' solution. Thus, the $G_3$ configuration is useful mostly to examine the impact phenomenon, appearance of atomization or the width of the impact zone which, as seen from Fig. \ref{synt_prof_th}b could be approximated, for $t=5.24\cdot 10^{-4}s$ at $d^*\approx3\cdot d$  which is consistent with the empirical formulation (\ref{nth4}) of \citep{elsaadawy}.

      To conclude the analysis of $G_3$, it is possible a new configuration could be proposed, similar to $G_3$ but involving periodic boundary conditions at the $y-$extends of the domain, gravity and a more varied initial thickness distribution possibly oscillating around Groenveld's thickness $h_G.$ Such a configuration would in turn allow one to mimic the physics described by (\ref{nth2}) or (\ref{nth6}) while keeping the computational cost low. However, obtaining the proper definition of boundary conditions for liquid entering and leaving the domain could be non-trivial; leaving it as an interesting option for a future research on this subject.

\section{Conclusions}

In this paper, we have presented a novel set of simulations of a very demanding, two-phase fluid flow whose characteristics closely correspond to that of air and liquid metal (Zinc). The boundary conditions correspond to the \textit{air knife} jet-wiping process in hot-dip coating. In many aspects this is a pioneering work: to our knowledge, the only similar calculations published have described a two-dimensional case with RANS/LES performed for the airflow  \citep{myrillas} or investigated the film formation only  \citep{snoe} or, possibly, included a predefined film (i.e. not formed gravitationally). Similarly, for reasons of numerical stability \citep{geo2015}, virtually all preceding attempts included a much decreased density ratio between the phases. Obviously, multitude of practical applications of similar results exist e.g. in metallurgical and automotive industries, however they are strictly proprietary and can not be consulted by general public.

None of these simplifications apply here: calculation accuracy for the methodology  presented here is limited only by available computational resources dictating the grid resolution. Full resolution of turbulent flow (i.e. below the Kolmogorov scale) is still too expensive.  However, thanks to grid adaptivity, we are able to achieve DNS in limited areas: this claim can be further justified considering the Hinze scale $l_H$, defined as the ratio of turbulent kinetic energy and surface tension, and estimated for the industrial parameters at $l_H\approx\num{1.76e-03}$m. At the grid resolution used in most cases presented here ($12$ levels of refinement) we obtain $l_H/\Delta x \approx 14,$ proving that $l_H$ is resolved. We can thus claim energy transfer from gas to liquid is at largely resolved, although of course we do not hold such claims for the turbulent flow in the gas itself.

Our results show that -- as expected in metal foundry practice -- the airflow inflicts a pressure gradient at the liquid layer, and ``punctuates'' it to a degree controlled by the $\Re$ in the air, and a properly defined Weber number. This gradient restricts the liquid feeding from reservoir, thinning the deposit. Our calculations fall short of resolving  the upper film thickness $h_c$ in the full $G_{1}$ and $G_2$ geometries, case $G_3$ being resolved the best, unfortunately the latter includes no gravitational flow. However, the $G_2$ case clearly displays the thinning effect, it is also observable in the $G_1$ case performed with decreased $\rho_l.$

We have observed levels of atomization of the liquid metal that were not previously reported in the literature. This phenomenon, to our knowledge, has also not been observed experimentally, which leads us to believe it is a purely transient effect, taking place only as a consequence of the initial gas-liquid impact event. It is predicted for $\We\approx 38$, while for $\We$ values closer to one, liquid wrinkling is observed, visible mostly in the $G_3$ configuration. The appearance of atomization seems thus predicted correctly, instead of being induced numerically. An additional observation is that the liquid material is ``milled'' (atomized) by the airknife before falling back to the reservoir, and that liquid-liquid collisions are aplenty. This is already visible at the $2^{12}$ level (Figures \ref{prof3}b and \ref{3dupdate}), and suggest that the coat-thinning mechanism is far more turbulent in its nature than known previously.

The three geometries introduced in the paper focus on the two-and three-dimensional coat formation and nozzle interactions ($G_1$), coat thinning and edge effects ($G_2$) and the character of gas-liquid impact ($G_3$). For future research -- apart from the aforementioned improvements to the $G_3$ configuration -- we would envision working preferable with the $G_2$ configurations in two and three dimensions, with increasing resolutions (and simulated time-spans), preferably until full resolution of the $h_c$ thickness is feasible.

\section*{Acknowledgements}

All the computations for this work have been performed using the French TGCC ``Irene Joliot-Curie'' and CINES ''Occigen'' supercomputers. Graphs and visualizations have been performed using the \textit{Basilisk View}\footnote{See \texttt{www.basilisk.fr/src/view.h}} Blender \citep{blender} and Gnuplot  \citep{gnuplot}.

The authors would like to thank Gretar Tryggvason for helpful discussions concerning the final state of this paper. We also thank Stefan Kramel, Stanley Ling, Sagar Pal, Maurice Rossi and Daniel Fuster for interesting discussions on the subject. We thank the participants of the ``DIPSI 2018'' workshop -- held in Bergamo and organized by G.E. Cossali and S. Tonini -- for their comments pertaining the early stages of this work. WA wishes to thank A.A.S. for proofreading the manuscript.

Wojciech Aniszewski would like to dedicate all his work on this paper to the dear memory of Antonina Gilska (nee Hessler, 1936-2019).

\bibliographystyle{elsarticle-harv}
\bibliography{ms}

\end{document}